\documentclass[appendixfloats]{aastex6}
\usepackage{graphicx}
\usepackage{natbib}
\usepackage{latexsym}
\usepackage{amssymb}
\usepackage{longtable}
\usepackage{amsmath}
\usepackage{url}
\def\msun{\ifmmode {\rm\,M_\odot}\else ${\rm\,M_\odot}$\fi}
\def\Msun{\ifmmode {\rm\,\it{M_\odot}}\else ${\rm\,M_\odot}$\fi}
\def\lsun{\ifmmode {\rm\,L_\odot}\else ${\rm\,L_\odot}$\fi}
\def\Lsun{\ifmmode {\rm\,\it{L_\odot}}\else ${\rm\,L_\odot}$\fi}
\def\rsun{\ifmmode {\rm\,R_\odot}\else ${\rm\,R_\odot}$\fi}
\def\Rsun{\ifmmode {\rm\,\it{R_\odot}}\else ${\rm\,R_\odot}$\fi}
\def\Tsun{\ifmmode {\rm\,T_\odot}\else ${\rm\,T_\odot}$\fi}
\def\arcsec{\ifmmode {^{\prime\prime}}\else $^{\prime\prime}$\fi}
\def\asec{\ifmmode {^{\prime\prime}}\else $^{\prime\prime}$\fi}
\def\arcmin{\ifmmode {^{\prime}}\else $^{\prime}$\fi}
\def\amin{\ifmmode {^{\prime}}\else $^{\prime}$\fi}
\def\simlt{\mathrel{\spose{\lower 3pt\hbox{$\mathchar"218$}}
     \raise 2.0pt\hbox{$\mathchar"13C$}}}
\def\simgt{\mathrel{\spose{\lower 3pt\hbox{$\mathchar"218$}}
\     \raise 2.0pt\hbox{$\mathchar"13E$}}}

\begin{document}

\title{A decade of H$\alpha$ transits for HD 189733 b: stellar activity versus absorption in the extended
atmosphere}

\author{P. Wilson Cauley and Seth Redfield}
\email{pcauley@wesleyan.edu}
\affil{Wesleyan University\\
Astronomy Department, Van Vleck Observatory, 96 Foss Hill Drive, Middletown, CT 06459}

\author{Adam G. Jensen}
\affil{University of Nebraska-Kearney\\
Department of Physics \& Astronomy, 24011 11th Avenue, Kearney, NE 68849}



\begin{abstract} 

HD 189733 b is one of the most well-studied exoplanets due to its large transit depth and host star
brightness.  The focus on this object has produced a number of high-cadence transit observations
using high-resolution optical spectrographs. Here we present an analysis of seven full H$\alpha$
transits of HD 189733 b using HARPS on the 3.6 meter La Silla telescope and HIRES on Keck I, taken
over the course of nine years from 2006 to 2015. H$\alpha$ transmission signals are analyzed as a
function of the stellar activity level, as measured using the normalized core flux of the
\ion{Ca}{2} H and K lines.  We find strong variations in the strength of the H$\alpha$ transmission
spectrum from epoch to epoch. However, there is no clear trend between the \ion{Ca}{2} core emission
and the strength of the in-transit H$\alpha$ signal, although the transit showing the largest absorption value
also occurs when the star is the most active. We present simulations of the in-transit contrast
effect and find that the planet must consistently transit active latitudes with very strong facular
and plage emission regions in order to reproduce the observed line strengths. We also investigate
the measured velocity centroids with models of planetary rotation and show that the small line
profile velocities could be due to large velocities in the upper atmosphere of the planet. Overall,
we find it more likely that the measured H$\alpha$ signals arise in the extended planetary atmosphere,
although a better understanding of active region emission for active stars such as HD 189733
are needed.

\end{abstract}


\section{INTRODUCTION}
\label{sec:intro}

Hot planets, or planets orbiting within $\sim$10 stellar radii of their host stars and which have
orbital periods of a few days, are unique subjects for planetary
astrophysics. The extreme proximity to their host stars can result in phenomena that are not
observable in other exoplanet systems \citep[for a recent review of these processes,
see][]{matsakos}. These phenomena include atmospheric mass loss
\citep[e.g.,][]{vidal,murray09,ehren12,bourrier13,ehren15,khodachenko,salz16}, magnetic and tidal star-planet interactions
\citep[e.g.,][]{cuntz,shkolnik08,lanza09,strugarek,poppenhaeger14,pillitteri15,miller15}, and bow
shocks that can form where the planet's atmosphere or magnetosphere plows through the stellar wind
\citep[e.g.,][]{lai,vidotto11,benjaffel,llama11,llama13,cauley15,turner16}. Due to the larger
magnitude of the potentially observable effect, giant hot planets ($M \gtrsim M_{Neptune}$) are
particularly good targets for investigating these processes. 

One of the most well studied hot Jupiters is HD 189733 b \citep{bouchy}. Due to the brightness of
its host star ($V=7.7$) and its large transit depth ($\sim$2.4\%), HD 189733 b has been the target
of many detailed observing campaigns over the last decade \citep[see Section 1.2 of][for a recent
overview]{bourrier13}. One of the most exciting results was the observation of evaporating material
by \citet{desetangs} who measured a 5\% transit depth at Lyman-$\alpha$ and concluded that the
absorbing material must be gravitationally unbound. A followup study showed that the evaporation is
highly variable: no absorption was detected in a set of 2010 Lyman-$\alpha$ observations while a
strong Lyman-$\alpha$ transit, with absorption up to $\sim$14\% measured in the blue wing of the
line profile, was detected in 2011 \citep{desetangs12}. These results were the first indication of
variability in the mass loss of a hot exoplanet. 

Evaporation of hot planets is driven by atmospheric heating from the absorption of X-ray and extreme
UV (EUV) stellar radiation \citep[e.g.,][]{murray09,owen12}. Two mechanisms can thus produce
variations in the amount of X-ray and EUV flux received by a hot planet: 1. stellar rotation and the
planet's orbital motion, which cause active regions\footnote{Throughout the paper we refer to active
regions as portions of the stellar surface covered by spots or faculae and plage.} of differing
strengths to be directed toward the planet throughout its orbit; and 2. intrinsic time variability
in the stellar activity level due to long-term activity cycles or short term variability, such as
flares. These variations provide a natural explanation for changes in the planetary evaporation rate
and suggest that hot planets subjected to larger amounts of ionizing stellar radiation will have
higher evaporation rates \citep{owen16}.

Measuring the exosphere of hot planets requires space-based UV observations
\citep[e.g.,][]{fossati,benjaffel,bourrier13}. However, the extended atmosphere, or the thermosphere
at pressures of 10$^{-6}$ - 10$^{-9}$ bar, can be observed from the ground using the neutral hydrogen
Balmer line transitions \citep{jensen12,christie,astudillo,cauley15,cauley16}. The first H$\alpha$
detection was made by \citet{jensen12}. \citet{christie} modeled the \citet{jensen12} detection and
showed that the strength of H$\alpha$ absorption in the atmosphere of a hot Jupiter is dependent on
the amount of ionizing stellar radiation, with larger amounts of radiation producing stronger
absorption. Thus while H$\alpha$ observations most likely do not directly probe the evaporating
material, the strength of absorption in the extended atmosphere may provide insight into the
strength of the evaporation. 

Recently, \citet{barnes16} used archival HARPS data to measure in-transit H$\alpha$ transmission
spectra for HD 189733 b across three separate transits. Based on velocity maps of the absorption
signatures in both the stellar and planetary rest frames, the authors point out that the absorption
signal moves from red-shifted to blue-shifted velocities in the planetary rest frame and shows no
velocity gradient across the transit in the stellar rest frame. They suggest that this is evidence of
the absorption signature arising in the frame of the star, i.e., it is not absorption by planetary
material but rather the result of continuum-to-line contrast effects that arise as the planet occults
different portions of the stellar disk. They also highlight the potential problems with using
continuum-to-line differential measurements for chromospherically sensitive lines: a non-uniform
stellar surface with active regions can be weighted towards a stronger or weaker
line core based on which portion of the stellar surface is occulted by the planet, resulting in
a contrast effect between the line core and nearby continuum region \citep{berta}. \citet{barnes16}
conclude that the transit signatures measured in the chromospherically sensitive lines, in particular
H$\alpha$ and the \ion{Ca}{2} H and K lines, are the result of the contrast effect and not
due to absorption by planetary material. This is especially relevant to active stars such as
HD 189733 \citep[$S_{HK}=0.52$; compared to the solar value of $S_{HK}=0.17$;][]{wright04}.   

In this paper we examine seven archival transits of HD 189733 b, including the same HARPS data
analyzed by \citet{barnes16}, in order to search for a relationship between the stellar activity
level and the strength of the in-transit H$\alpha$ signal. We do not include the results of
\citet{jensen12} since these observations were not performed across a single transit.  We produce
detailed transit models in order to investigate the contrast effect and determine if the H$\alpha$
signatures measured in \citet{jensen12}, \citet{cauley15}, \citet{cauley16}, and \citet{barnes16}
can be attributed to occultation of a non-uniform stellar surface rather than absorption by
planetary material. We also discuss how planetary rotation can affect the measured absorption
velocities, which we calculate for each transit, and present transmission spectrum models of
rotation in an extended planetary atmosphere. The investigation into the absorption line
velocities is motivated by the suggestion of \citet{barnes16} that the H$\alpha$ transmission
is not due to the planetary atmosphere. Their main argument is that the H$\alpha$ line
velocities do not follow the pattern expected from absorption in the planetary atmosphere,
although we note that they do not present quantitative measurements of the line velocities.
The data sets and reduction processes are
described in \autoref{sec:observations}. The transmission spectrum is defined in
\autoref{sec:transspec} and the average H$\alpha$ transmission spectra are given.  H$\alpha$ and
\ion{Ca}{2} H and K time-series measurements are presented in \autoref{sec:timeseries} for each
transit. \autoref{sec:epochs} includes an examination of epoch to epoch changes in H$\alpha$
absorption and the stellar activity level as measured using the \ion{Ca}{2} H and K lines. The
contrast models are discussed in \autoref{sec:contrast} and the atmospheric rotation models and
measured line velocities are given in \autoref{sec:velocities}. \autoref{sec:conclusion} provides a
brief summary and conclusion of our results.

\section{Observations and data reduction}
\label{sec:observations}

The observations presented here are a combination of HD 189733 b transits observed with HARPS
\citep{mayor03} on the 3.6-meter telescope at La Silla and HIRES \citep{vogt} on Keck I. Information
about each data set is detailed in \autoref{tab:tab1}, including the average signal-to-noise (S/N) of
an individual exposure in the continuum near 6571 \AA. All data is currently available on either the
ESO data archive or the Keck Observatory Archive. The archive ID for each data set is given in
\autoref{tab:tab1}. We note that the HARPS data is the same used by \citet{wyttenbach} and \citet{louden} to study
\ion{Na}{1} absorption in HD 189733 b's atmosphere. The Keck data from 2006 August 21 is the same
data used to measure the Rossiter-McLaughlin effect for HD 189733 b by \citet{winn06}. We also 
retrieved an archival transit obtained using UVES \citep{dekker} by \citet{czesla15} to study center-to-limb
variations of the \ion{Na}{1} D lines. This transit, however, is fundamentally different when compared to all of the
other H$\alpha$ transits and does not match any of the features seen in the HIRES and HARPS
time series. This abnormal behavior is due to the mid-transit flare identified by \citet{czesla15} and
further investigated by \citet{klocova17}. Finally, we exclude archival transits from the High Dispersion Spectrograph
on Subaru due to the lack of simultaneous \ion{Ca}{2} observations.    

HARPS has a resolving power of $R \sim 115,000$ and the HIRES observations were
performed at $R \sim 49,000$ using the B5 decker and $R \sim 68,000$ using the B2 decker. The reduced
HARPS data were taken directly from the ESO archive. We note that we do not perform the scattered
light removal process described by \citet{barnes16}. We believe that a comparison of our timeseries
results and average transmission spectra provide justification: we obtain almost identical
results for the behavior of both H$\alpha$ and \ion{Ca}{2} using the standard HARPS reduction routines. 

Standard reduction steps including bias subtraction, flat fielding, and wavelength calibration 
were performed for the Keck data using the publicly available HIRES Redux program by Jason X.
Prochaska\footnote{http://www.ucolick.org/$\sim$xavier/HIRedux/}. All spectra are shifted to the
rest frame of the star by correcting for the Earth's barycentric velocity and HD 189733's
system radial velocity, for which we use the mean value $-2.23$ km s$^{-1}$ from \citet{digloria15}.

We used the latest version of Molecfit \citep{kausch} in order to model telluric absorption in the
H$\alpha$ order. We first construct a master telluric model by using a telluric standard observed on
the same night. The master telluric model is then fit to a selection of telluric lines in the
individual science exposures using a $\chi^2$ minimization routine. We fit for the line depth, small
wavelength shifts, and line broadening. The best-fit scaled, broadened, and shifted telluric model
is then divided out of the normalized science spectrum. This routine works very well for most
spectra, removing the telluric absorption down to 5 - 10\% of the original line depth. This typically
results in transmission spectrum residuals of $\sim$0.1 - 0.5\% near the cores of the telluric lines.
Due to the low signal-to-noise, we were unable to adequately remove the telluric absorption
immediately redward of H$\alpha$ for the 2007 August 29 observations (see \autoref{fig:hatransits}).
This portion of the spectrum is not included in the absorption calculations. 

Two-element wavelength binning is applied to all of the individual HARPS spectra in order to
increase the signal-to-noise (S/N). We also co-add back-to-back spectra from the nights of
2007 July 20 and 2007 August 29 due to the short exposure times (300 seconds). Note that the S/N values
for the HARPS spectra in \autoref{tab:tab1} are calculated for the co-added and binned spectra.  The
individual Keck observations required no binning. Some spectra near the beginning or end of the
night are not included in the analysis due to $S/N \lesssim 50$ or contamination by Earth's twilight
sky spectrum. 

\begin{deluxetable}{lccccccc}
\tablecaption{Archived data sets \label{tab:tab1}}
\tablehead{\colhead{UT Date}&\colhead{Instrument}&\colhead{Program ID}&\colhead{$N_{obs}$}&
\colhead{$N_{used}$}&\colhead{S/N @ 6571 \AA}&\colhead{\textit{R}}&\colhead{$\lambda_{start}$,$\lambda_{end}$}}
\colnumbers
\tabletypesize{\scriptsize}
\startdata
2006 Aug 21 & Keck HIRES & A259Hr & 70 & 55 & 230 & 49,000 & 3360\AA,7800\AA \\
2006 Sep 8 & HARPS & 072.C-0488(E) & 18 & 18 & 149 & 115,000 & 3800\AA,6800\AA \\
2007 Jul 20 & HARPS & 079.C-0828(A) & 39 & 39 & 140 & 115,000 & 3800\AA,6800\AA\\
2007 Aug 29 & HARPS & 079.C-0127(A) & 40 & 40 & 131 & 115,000 & 3800\AA,6800\AA\\
2013 Jun 3 & Keck HIRES & A308Hr & 17 & 16 & 440 & 68,000 & 3500\AA,7400\AA \\
2013 Jul 4 & Keck HIRES & A308Hr & 40 & 37 & 490 & 68,000 & 3500\AA,7400\AA \\
2015 Aug 4 & Keck HIRES & N120Hr & 61 & 61 & 490 & 68,000 & 3500\AA,7400\AA \\
\enddata
\end{deluxetable}

\section{Transmission spectra}
\label{sec:transspec}

The transmission spectrum is defined here as:

\begin{equation}\label{eq:strans}
S_T=\frac{F_{i}}{F_{out}}-1
\end{equation}

\noindent where $F_i$ is a single in-transit observation and $F_{out}$ is the master comparison
spectrum. To produce the final transmission spectra, we apply the same wavelength alignment and
normalization procedures described in \citet{cauley15,cauley16}. The comparison spectra are
generally chosen to be those taken furthest from the transit, although there is no strict rule for
doing so, and their number is selected to minimize the impact of any single spectrum while not using
too many of the available observations as comparison exposures. For example, the 2006 August 21
comparison spectra are chosen as those immediately before the transit in order to balance the low
$W_{H\alpha}$ values near $t-t_{mid}=-100$ minutes and the increase in $W_{H\alpha}$ immediately
after the transit begins. The choice of comparison spectra can significantly influence the absolute
level of measured absorption but the relative changes between observations remain the same. For this
study, the choice of out-of-transit comparison spectra does not significantly change the results for
any of the transits except the 2015 Aug 4 data. The choice of comparison spectra for 2015 Aug 4 is
detailed in \citet{cauley16}. 

\begin{figure*}[htbp]
   \centering
   \includegraphics[scale=.85,clip,trim=5mm 10mm 5mm 0mm,angle=0]{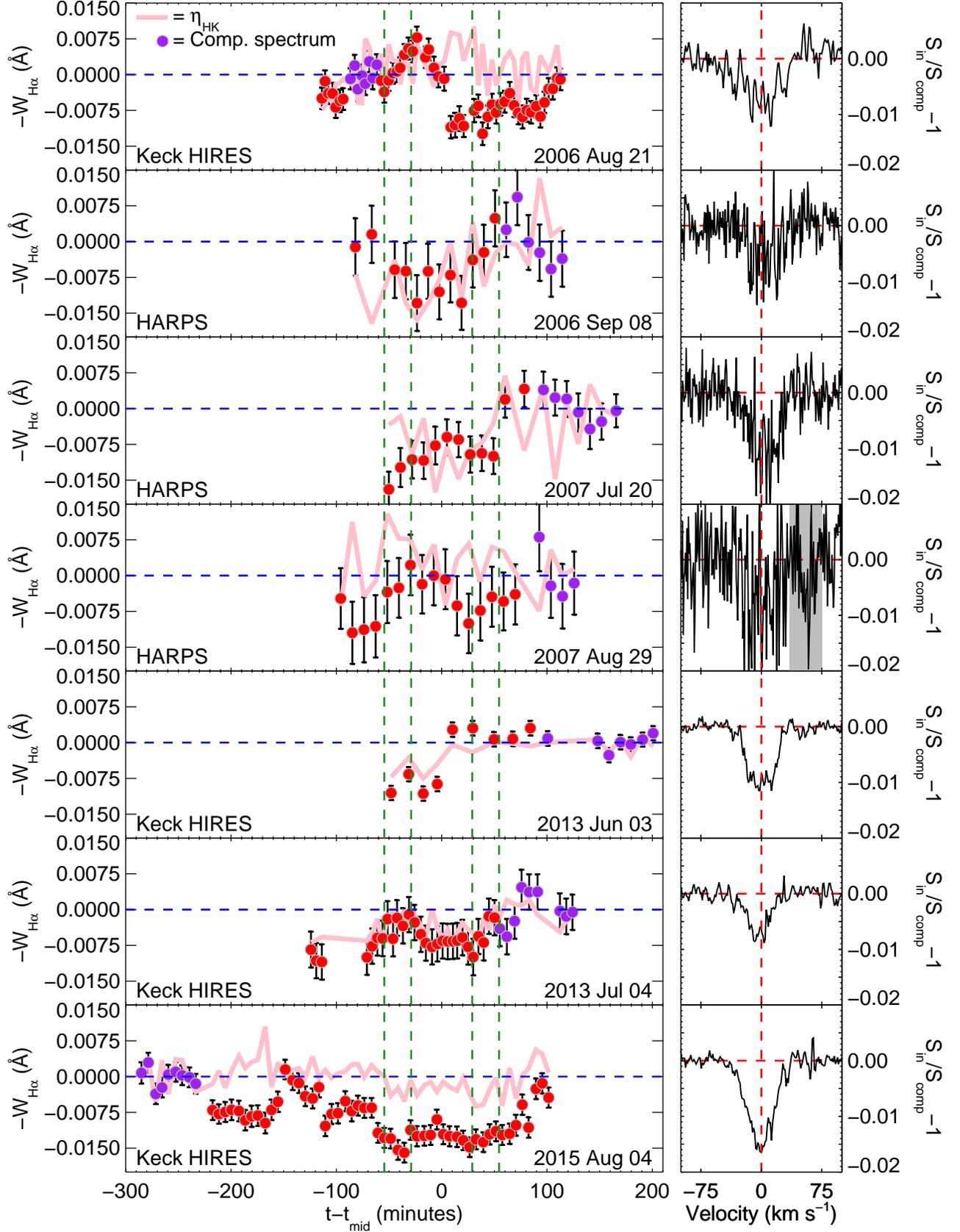} 
   \figcaption{H$\alpha$ absorption timeseries for the seven dates from \autoref{tab:tab1}. The
$W_{H\alpha}$ points are shown with red circles. Observations used as comparison spectra are shown
as purple circles. All dates are shown on the same scale. The thick pink line shows the scaled and
median corrected $\eta_{HK}$ values in order to convey how the \ion{Ca}{2} flux changes correlate
with changes in $W_{H\alpha}$. The right column shows the master
in-transit transmission spectrum for each transit. Note that only observations showing 1$\sigma$
absorption are included in the master transmission spectrum.\label{fig:hatransits}}
\end{figure*}

The average H$\alpha$ in-transit transmission spectrum for the individual transits is shown in the right side
column of \autoref{fig:hatransits}. Only individual in-transit observations showing 1$\sigma$
significant absorption are selected to be included in the average transmission spectra in order to
highlight the line morphology. Including all of the in-transit observations from 2013 June 3, for
example, results in a much weaker transmission spectrum due to the abrupt decrease in absorption
mid-transit. Master absorption measurements (see \autoref{eq:wlambda}) are given for each date in
\autoref{tab:tab2}. Absorption is detected at the 3$\sigma$ level for all transits. 

Significant variations in $S_T$ from epoch to epoch are evident, especially in the high S/N Keck
spectra. The HARPS spectra are fairly noisy but clear changes in line depth, and even shape, can be
seen. These variations suggest that if the signal arises in HD 189733 b's extended planetary
atmosphere, it changes rather drastically from one epoch to the next, and perhaps within individual
epochs, as observed in the Ly$\alpha$ exosphere by \citet{desetangs} and \citet{desetangs12} and
across the multiple H$\alpha$ transits observed by \citet{jensen12}.

\section{Time-series H$\alpha$ absorption and the normalized \ion{Ca}{2} core flux}
\label{sec:timeseries}

For each individual spectrum we calculate the following absorption measure, essentially an equivalent width
of the transmission spectrum, at H$\alpha$:

\begin{equation}\label{eq:wlambda}
W_{H\alpha} = \sum\limits_{v=-200}^{+200} \left(1-\frac{F_v}{F_v^{out}} \right) \Delta\lambda_v 
\end{equation}

\noindent where $F_v$ is the flux in the spectrum of interest at velocity $v$, $F_v^{out}$ is the flux in the
comparison spectrum at velocity $v$, and $\Delta\lambda_v$ is the wavelength difference at velocity
$v$. The units of $W_{H\alpha}$ are angstroms. The gray shaded region in the transmission spectrum 
for 2007 August 29 is ignored due to poor telluric subtraction. 

Individual HARPS transmission spectra are normalized across the $\pm 200$ km s$^{-1}$ region by
averaging the fits of a line and a low order spline. The $\pm 40$ km s$^{-1}$ at line center are 
ignored in the normalization. Individual Keck spectra, which are much higher S/N than the HARPS
transmission spectra and have fewer telluric residuals, require only a second-order polynomial to
adequately remove the continuum slope. We note that small residual offsets from zero in the normalized HARPS transmission spectra
can result in $W_{H\alpha}$ offsets of $\sim$0.001 \AA. These residuals, however, are also present in
the $W_{H\alpha}$ values calculated for the individual comparison points. For this reason, we do not
separately include the normalization uncertainties.

We derive uncertainties in $W_{H\alpha}$ by combining in quadrature two different sources
of uncertainty. First, normalized flux errors in the transmission spectrum are summed in quadrature. This
is then added in quadrature to the standard deviation of the comparison spectra $W_{H\alpha}$ points
(purple circles in \autoref{fig:hatransits}). The standard deviation uncertainties dominate
the normalized flux uncertainties in most cases. We note that this is different from the empirical Monte Carlo (EMC)
procedure used in \citet{redfield}, \citet{jensen12}, \citet{cauley15}, and \citet{cauley16} but has a similar outcome: large variations in the comparison
spectra, like in the 2013 July 4 transit, will produce larger uncertainties in all of the individual points. We
choose to use the standard deviation of the comparison points since the HARPS time series, compared with
the 2013 and 2015 Keck time series, have relatively fewer comparison spectra ($N=4-7$ compared to $N=8$ for 
the 2013 and 2015 Keck nights). The EMC procedure 
becomes less useful with smaller numbers of comparison spectra. The standard deviation uncertainties are
$\sim$2.6 times greater than the EMC uncertainties for the individual 2013 and 2015 Keck data points. This
is approximately equal to $\sqrt{N}$ where $N$ is the number of comparison spectra used. We recommend
that future studies adopt the more conservative standard deviation estimate for individual time series points.

\begin{deluxetable*}{lccc}
\tablecaption{H$\alpha$ absorption and $\eta_{HK}$ values\label{tab:tab2}}
\tablehead{\colhead{}&\colhead{$W_{H\alpha}^a$}&\colhead{Mean in-transit $-\eta_{HK}^b$}&\colhead{Mean comparison $-\eta_{HK}$}\\
\colhead{UT Date}&\colhead{(10$^{-3}$ \AA)}&\colhead{(\AA)}&\colhead{(\AA)}}
\colnumbers
\tabletypesize{\scriptsize}
\startdata
2006 Aug 21 & 9.11$\pm$0.94 & 0.373$\pm$0.019 & 0.370$\pm$0.025  \\
2006 Sep 8 & 9.29$\pm$1.10 & 0.358$\pm$0.046 & 0.424$\pm$0.043  \\
2007 Jul 20 & 10.00$\pm$1.07 & 0.246$\pm$0.043 & 0.291$\pm$0.056  \\
2007 Aug 29 & 7.89$\pm$2.30 & 0.299$\pm$0.050 & 0.245$\pm$0.075  \\
2013 Jun 3 & 9.14$\pm$0.32 & 0.389$\pm$0.011 & 0.415$\pm$0.008  \\
2013 Jul 4 & 6.84$\pm$0.42 & 0.346$\pm$0.014 & 0.364$\pm$0.013  \\
2015 Aug 4 & 12.80$\pm$0.45 & 0.540$\pm$0.013 & 0.558$\pm$0.025  \\
\enddata
\tablenotetext{a}{Uncertainties are calculated by propagating the flux uncertainty for each spectrum through \autoref{eq:wlambda}.}
\tablenotetext{b}{Uncertainties are the standard deviation of the points included in the mean.}  
\end{deluxetable*}

\begin{figure*}[t]
   \centering
   \includegraphics[scale=.72,clip,trim=7mm 40mm 15mm 50mm,angle=0]{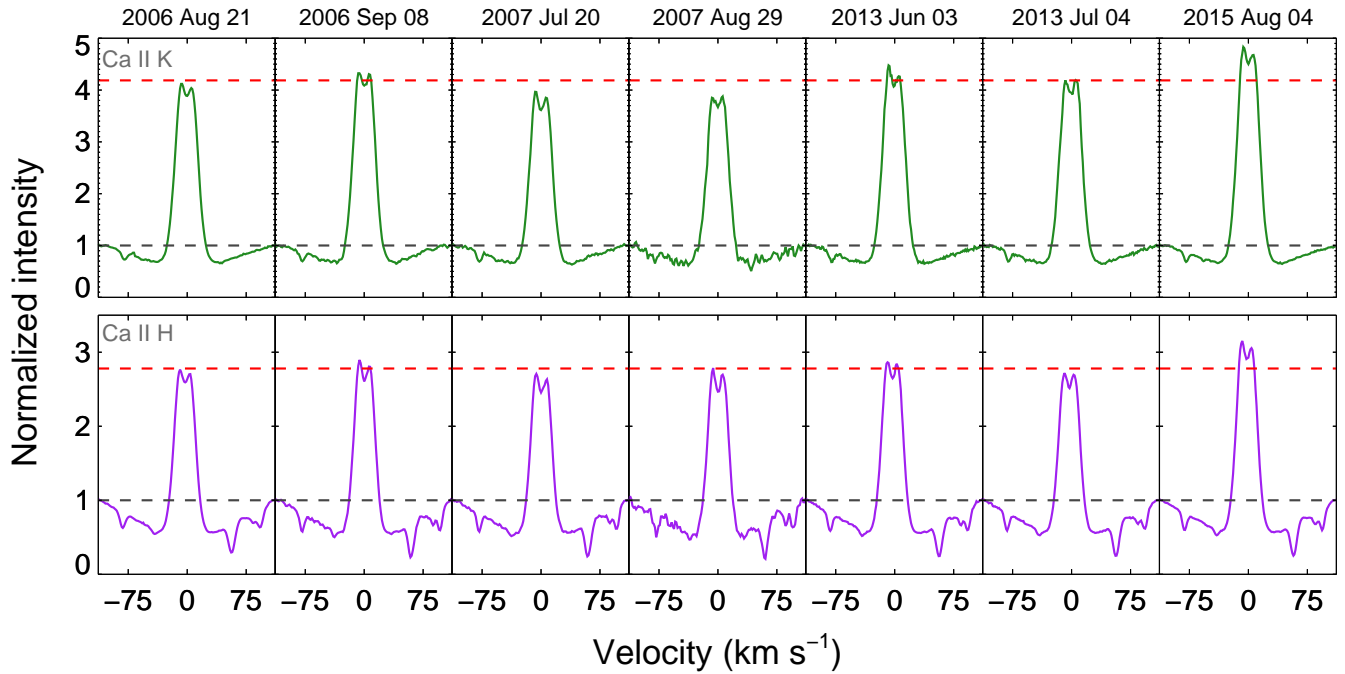} 
   \figcaption{The normalized mean \ion{Ca}{2} H and K profiles for the comparison spectra from each date. The dashed
charcoal line shows the normalization level and the dashed red line shows the median maximum flux
level for all dates. Note the strong excess core emission for 2015 August 4. \label{fig:caiiprofs}}
\end{figure*}

In order to test for correlations between the stellar activity level and $W_{H\alpha}$, we define the
following measure of the average between the \ion{Ca}{2} H and K core fluxes:

\begin{equation}\label{eq:etahk}
\eta_{HK} = \frac{1}{2} \left[ \sum\limits_{}^{} \left( F_N^H- 1\right) \Delta\lambda_H + \sum\limits_{}^{} \left( F_N^K - 1\right) \Delta\lambda_K \right]
\end{equation}

\noindent where $F_N^H$ is the core flux of the \ion{Ca}{2} H line normalized to the 0.1 \AA\ wide
regions centered at $\pm$1.5 \AA\ from the rest wavelength of the line. $\Delta\lambda_H$ is the
dispersion near \ion{Ca}{2} H and the flux integration is done between $\lambda_0^H \pm
1.5$ \AA. The symbols have the same meaning for \ion{Ca}{2} K. We note that we do not perform the
residual \ion{Ca}{2} H and K core analysis from \citet{cauley16} due to the low S/N of the HARPS
spectrum. The $\eta_{HK}$ measures essentially the same thing but information about the shape of
the residual line profile is lost. In-transit and comparison spectra $\eta_{HK}$ values are given in
\autoref{tab:tab2}. The mean \ion{Ca}{2} H and K profiles for the comparison observations are shown
in \autoref{fig:caiiprofs}. The median maximum normalized flux is marked with a red dashed line. 

\autoref{fig:hatransits} shows $W_{H\alpha}$ (red circles) as a function of the time from
mid-transit for each of the transits listed in \autoref{tab:tab1}. All rows are on the same vertical
and horizontal scale. The thick pink lines show a scaled value of $\eta_{HK}$ that has been shifted
by the median $\eta_{HK}$ value of the comparison spectra (purple circles). The vertical green lines
show the transit contact points and the horizontal blue line marks $W_{H\alpha}=0.0$. Consistent
transits in H$\alpha$ are detected for the HARPS data on 2006 September 8 and 2007 July 20.
Transits showing transient absorption signatures are detected for 2006 August 21, 2007 August 29,
and 2013 June 3. The shape of the $W_{H\alpha}$ HARPS timeseries measurements are very similar to
those presented by \citet{barnes16}.

A few things are immediately evident from \autoref{fig:hatransits}. First, the H$\alpha$ signal is
highly variable, showing large deviations from epoch to epoch in both the strength of the absorption
and in the duration of the transit. Second, the data from 2006 August 21, 2007 August 29, and 2013
June 3 show abrupt in-transit changes in $W_{H\alpha}$. This contrasts with the other four dates
which show fairly uniform in-transit absorption that is typical of, for example, a non-varying
extended atmosphere. Lastly, 2006 August 21, 2013 July 4, and 2015 August 4 show evidence of
absorption outside of the optical transit times: 2006 August 21 shows post-transit absorption, 2013
July 4 shows pre-transit absorption, and 2015 August 4 shows both pre- and post-transit absorption.
The pre-transit signals from 2013 July 4 and 2015 August 4 have been explored in
\citet{cauley15,cauley16}. The abrupt in-transit changes could be due to transits of active regions on the
stellar surface or varying levels of stellar activity. Active region transits will be investigated in
\autoref{sec:contrast} and correlations with the \ion{Ca}{2} H and K lines will be discussed 
in the next subsection.

\begin{figure}[htbp]
   \centering
   \includegraphics[scale=.87,clip,trim=65mm 12mm 50mm 0mm,angle=0]{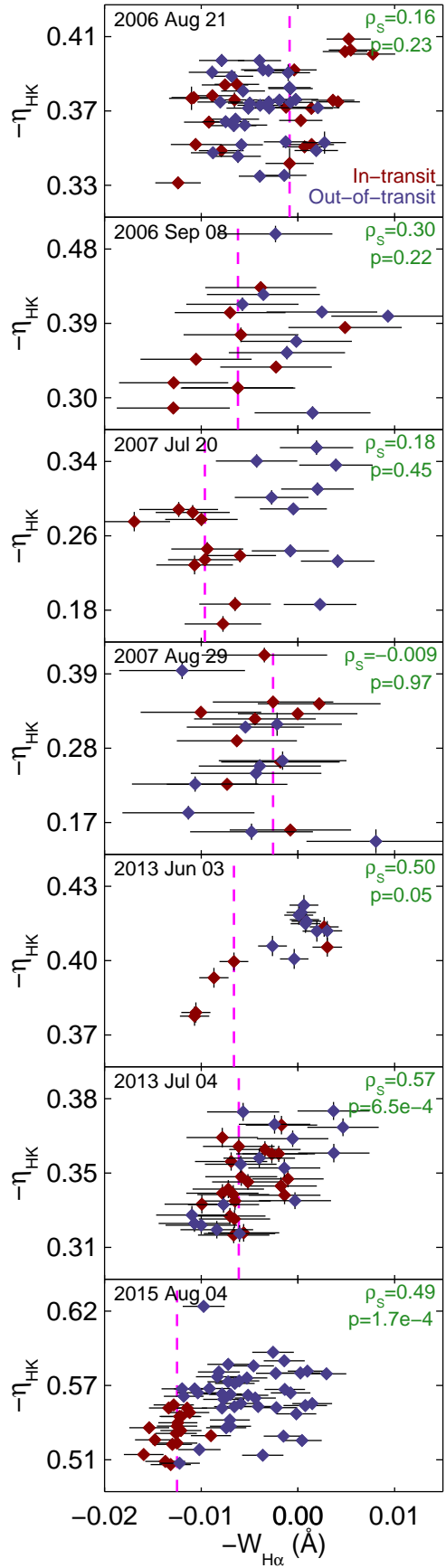} 
   \figcaption{Correlation plots between $W_{H\alpha}$ and $\eta_{HK}$. In-transit points are shown
in dark red and out-of-transit points are shown in dark blue. The median in-transit $W_{H\alpha}$
value is marked with a vertical magenta dashed line. Correlations significant at the 99\%
level are present for 2007 July 20 and 2013 July 4. \label{fig:corrplots}}
\end{figure}

\subsection{Stellar activity versus planetary absorption}

As discussed in \citet{cauley15,cauley16}, changes in the stellar activity from one observation to
another can mimic absorption by circumplanetary material. We note that this is separate from the
in-transit contrast effect, which is discussed in \autoref{sec:contrast}. One method of distinguishing absorption from
stellar variability is to compare the $W_{H\alpha}$ values with simultaneous measurements of an independent measure
of the stellar activity level. Changes in $W_{H\alpha}$ can be more confidently attributed to absorption
if there is no correlation with the independent stellar activity measure. 

\autoref{fig:corrplots} shows the values of $W_{H\alpha}$ versus $\eta_{HK}$ for each epoch. In-transit
observations are shown in dark red while out-of-transit observations are shown in dark blue. We calculate
Spearman's $\rho$ rank correlation coefficients for each date. The $\rho_S$ value and corresponding
two-sided significance $p$ are shown in green in the upper-right of each panel. Only the dates of
2013 June 3, 2013 July 4, and 2015 Aug 4 show significant ($p\lesssim0.05$) correlations between $W_{H\alpha}$ and $\eta_{HK}$.
The 2015 Aug 4 correlation is largely driven by the distinct in- and out-of-transit groupings, i.e.,
there is no correlation \textit{within} the in-transit points or within the out-of-transit points.
The clumping of the points from 2013 June 3 likely prevents the correlation from being stronger but
$\eta_{HK}$ clearly changes at the same time as $W_{H\alpha}$ near $t-t_{mid}=0$ minutes. 
We discuss the relationship between $W_{H\alpha}$ and $\eta_{HK}$ for the uniform $W_{H\alpha}$
transits in the next section.

For the non-uniform transits of 2006 August 21, 2007 August 29, and 2013 June 3, the interpretation of the
$\eta_{HK}$ and $W_{H\alpha}$ relationship is uncertain. The small number of comparison points
and low S/N of the 2007 August 29 data make this transit especially difficult to interpret. The 2006 August 21 
and 2013 June 3 transits both feature abrupt changes in $W_{H\alpha}$ at mid-transit, 
producing strong absorption in the case of 2006 August 21 and filling in the
absorption in the case of 2013 June 3. The clear change in $\eta_{HK}$ at a similar time in the
2013 June 3 transit suggests that this may be entirely attributable to changes in the stellar 
activity level. This is not the case for 2006 August 21.

It seems unlikely that such abrupt changes can be due to physical variations in the planetary atmosphere. 
However, transiting gas (e.g., previously evaporated material) not associated with the atmosphere
could cause such changes. In this case, the atmosphere would show no absorption and the
external feature \citep[e.g., a condensation or accretion stream;][]{lai,lanza14}
 would be orbiting ahead of the planet in the case of 2013 June 3 and behind
the planet for the 2006 August 21 data. The feature would have to be $\sim$5 $R_p$ ahead of or behind
HD 189733 b for the transit to end or begin halfway through the planet's transit. 
Although it is unclear as to the exact nature of these short-term $W_{H\alpha}$ variations, which
are present in all of the transits, we showed in \citet{cauley17} that changes of
the magnitudes seen, for example, in the 2013 June 3 and 2006 August 21 data, 
are rare when the planet is out-of-transit, suggesting that they occur preferentially
when the planet is in- or near-transit. We find it more likely that these changes are due 
variable absorption in the circumplanetary environment since stochastic changes in the stellar activity level,
as a result of star-planet interactions, should be observable at orbital phases not associated
with the transit.

\subsection{\ion{Ca}{2} as an absorber}

Although \ion{Ca}{2} emission is a widely examined indicator of chromospheric activity, it is possible that \ion{Ca}{2}
atoms in the extended atmosphere of HD 189733 b might have enough opacity to absorb stellar photons \citep{turner16}.
\citet{lanza14} posited that \ion{Ca}{2} absorption in stellar prominences fed by planetary mass loss
could be responsible for the correlation between planetary surface gravity and log$R^{'}_{HK}$ found
by \citet{hartman10} \citep[see also][]{fossati15}. 
Indeed, detections of multiple transitions of neutral calcium have been claimed in the atmosphere of HD 209458 b \citep{astudillo}. If
there is significant \ion{Ca}{2} absorption by the planet, the measured in-transit core flux is no longer directly
tracing the short-term stellar activity level, especially when comparing in-transit observations to out-of-transit
observations. This is hinted at by the fact that 5 of 7 transits have lower mean in-transit $\eta_{HK}$ values compared to the
comparison spectrum $\eta_{HK}$ value (see \autoref{tab:tab2}). We note, however, that the probability of
measuring a decrease in $\eta_{HK}$ in at least 5 of the 7 transits is $\sim$22\%, assuming that there is
an equal probability of measuring either an increase or decrease. Thus we cannot say with any statistical
certainty that the in-transit \ion{Ca}{2} measurements show a preference for absorption. Increasing the
number of transits at \ion{Ca}{2} would help clarify if the observed statistics are representative of a real effect.

Planetary \ion{Ca}{2} absorption may explain the data for the dates showing lower in-transit $-\eta_{HK}$ values and fairly
smooth $W_{H\alpha}$ transits. For the 2006 September 8, 2007 July 20,
2013 July 4, and 2015 August 4 transits the mean $\eta_{HK}$ value for the comparison spectra is more negative (meaning more core flux)
than the mean in-transit $\eta_{HK}$ value. We find significant correlations between $W_{H\alpha}$
and $\eta_{HK}$ for 2013 July 4 and 2015 Aug 4. The 2013
July 4 correlation is driven by a correlation \textit{within} the out-of-transit points ($\rho_S=0.71$, $p=0.002$); none of
the other dates show correlations within only the in- or out-of-transit points. However, the 2013 July 4
out-of-transit correlation is itself driven by strong differences between the pre-transit points and
the post-transit comparison points (see \autoref{fig:hatransits}). Thus we do not see strong evidence of correlated changes 
between $W_{H\alpha}$ and $\eta_{HK}$ from one exposure to another. This tentatively suggests that the lower in-transit
$\eta_{HK}$ differences are not due to changes in the stellar activity level.


An alternative to absorption in the planetary atmosphere producing the lower in-transit $\eta_{HK}$
values is the transiting of bright active regions \citep{llama15}. If the planet happens to transit
an especially active latitudinal band that is bright in \ion{Ca}{2} compared to the rest of the star,
the \ion{Ca}{2} core emission will be preferentially blocked by the planet's disk, resulting in lower
in-transit $\eta_{HK}$ values. Unless the active latitude is fairly uniform in brightness, the transit tends to
be choppy and uneven \citep{llama15}. Although the \ion{Ca}{2} light curves in \autoref{fig:hatransits} do
show some significant exposure-to-exposure variations, the epochs with lower in-transit versus out-of-transit
$\eta_{HK}$ values (e.g., 2007 July 20 and 2013 July 4) exhibit fairly smooth time series. Thus we
tentatively favor the atmospheric absorption scenario over active region transits, although a
thorough investigation of the contrast effect at \ion{Ca}{2} H and K would be informative. 


If \ion{Ca}{2} in the planetary atmosphere is indeed absorbing, it cannot be considered a reliable
independent measure of the stellar activity level near (i.e., immediately pre- or post-transit) 
or during the planet's transit. Detailed theoretical modeling of the \ion{Ca}{2} population in the 
extended atmospheres of hot exoplanets would be useful in determining if \ion{Ca}{2} 
should be specifically targeted in these systems. A possible alternative is simultaneous monitoring
of a FUV activity diagnostics \citep[e.g.,][]{france16} that are not expected to be present in
the extended atmospheres of hot planets. 

\section{Epoch to epoch variations}
\label{sec:epochs}

The significant exospheric changes measured in Ly$\alpha$ by \citet{desetangs12} and \citet{bourrier13} for HD 189733 b motivates
us to look for changes in the H$\alpha$ signal as a function of stellar activity level. This has also been
investigated previously by \citet{barnes16} for the HARPS transits. \citet{christie} showed
that the strength of H$\alpha$ planetary absorption depends strongly on the Ly$\alpha$ and ionizing flux from the star. In order
to match the strength of H$\alpha$ absorption measured by \citet{jensen12}, \citet{christie} find that a large
value of the ionizing flux is needed. For reference, the H$\alpha$ absorption measured by \citet{jensen12} is
comparable to that measured for the 2015 August 4 transit. This result suggests that dates showing larger
values of out-of-transit $\eta_{HK}$ should show larger amounts of in-transit H$\alpha$ absorption.

\begin{figure}[t]
   \centering
   \includegraphics[scale=.53,clip,trim=52mm 26mm 25mm 35mm,angle=0]{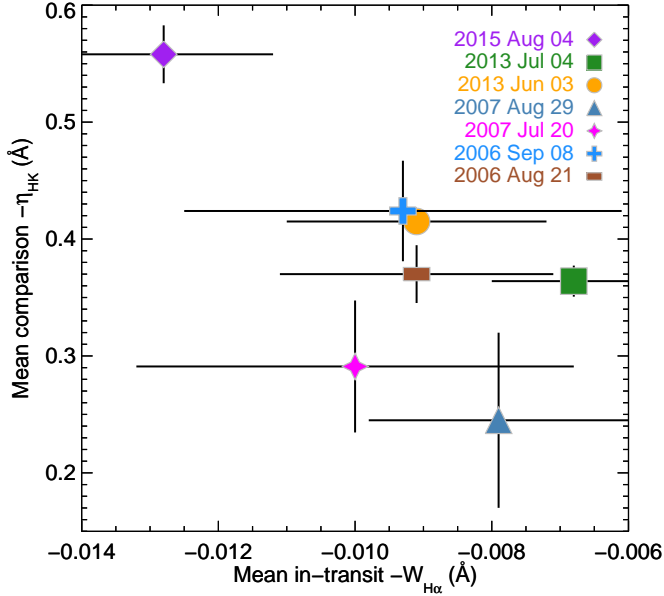} 
   \figcaption{Comparison between the mean in-transit values of $W_{H\alpha}$ and $-\eta_{HK}$ calculated
   for the mean \ion{Ca}{2} H and K comparison spectra shown in \autoref{fig:caiiprofs}. 
   Uncertainties in $\eta_{HK}$ for 2013 June 3 are smaller than the size of the 
   plot symbol. There is no clear trend. However, the date with the highest $\eta_{HK}$ by a large margin is 2015 August 4
(purple) and this date also shows the deepest and most extended H$\alpha$ transit, suggesting that the
extended atmosphere is strongly influenced by the stellar activity level if the H$\alpha$
signal arises in the planetary atmosphere. \label{fig:havca}}
\end{figure}

The strongest core emission in \autoref{fig:caiiprofs} is clearly seen for 2015 August 4. 
\autoref{fig:havca} shows the mean $-\eta_{HK}$ value from
\autoref{fig:caiiprofs} plotted against the mean in-transit $-W_{H\alpha}$ from each date. The 
uncertainty in each value is the standard deviation of the points included in the mean value. Only points showing
1$\sigma$ significant absorption are included in the mean $W_{H\alpha}$ values. 

There is no clear trend between $W_{H\alpha}$ and $\eta_{HK}$ from epoch to epoch. However, 
2015 August 4 stands out: it shows the strongest in-transit H$\alpha$ absorption signal and the largest
out-of-transit $\eta_{HK}$ value. In addition, the 2015 August 4 H$\alpha$ transit appears significantly extended
both pre- and post-transit, hinting at an atmosphere that fills a significant fraction of the planet's
Roche lobe. The low H$\beta$ and H$\gamma$ absorption values from \citet{cauley16} also
support the presence of a low-density extended atmosphere. This suggests that there may be 
some threshold (e.g., $-\eta_{HK}$ $\gtrsim$ 0.40)
for the stellar activity level below which the extended atmosphere is fairly uniform and constant.
We note that the in- versus out-of-transit differences in $\eta_{HK}$ within a single night
are much less than the inter-epoch variations. Thus although we have argued that \ion{Ca}{2} 
may be absorbing stellar photons around the planet, we do not expect the absorption effect
to be as large as night to night variations in the stellar activity level, as seen, for example, by \citet{boisse09}
across many weeks of H$\alpha$ and \ion{Ca}{2} observations of HD 189733 and similarly at H$\alpha$ by \citet{cauley17}.

It is not clear if the star was in a particularly active state during the 2015 August 4 transit or an 
especially active hemisphere of the star was visible during that night. 
The suggested threshold needs to be confirmed with a larger sample of transits, ideally during periods when the star
is in an especially active state, that show consistent H$\alpha$ absorption since the interpretation 
of irregular transits (e.g., 2006 August 21 and 2007 August 29) is much less straightforward.  

Overall, we find no strong evidence for a relationship between the stellar activity level
and the strength of the in-transit $W_{H\alpha}$ measurement, although the date with
the highest $\eta_{HK}$ value also shows the strongest in-transit H$\alpha$ absorption.
The analysis presented here should be supplemented in the future with more transit
observations, preferably containing as many simultaneous activity indicators as possible.
In the next section, we investigate whether or not the contrast effect for HD 189733 can
explain the observed H$\alpha$ absorption signatures. A similar analysis will be presented
in a future paper for \ion{Ca}{2}, \ion{Na}{1}, and \ion{Mg}{1}.

\section{The contrast effect for HD 189733}
\label{sec:contrast}

While much effort has been made to calculate the effects of star spots, faculae and plage\footnote{We do not distinguish
between faculae, which form in the photosphere, and plages, which form in the chromosphere. The spectra
used to model bright H$\alpha$ regions on the stellar surface are a combination of both plage and facular
characteristics.}, and other irregular surface features on stellar radial velocities, broadband transmission spectra, and 
properties derived from transit measurements 
\citep[e.g., ][]{saar97,aigrain12,dumusque14,oshagh14,andersen15,llama15,herrero16,giguere16,rackham17},
little has been published on the details of how these same features affect high-resolution transmission spectra.
\citet{berta} and \citet{sing11} present analytic approximations for how the broadband transmission
spectrum changes as a function of star spot filling factor and temperature. Here we
present an investigation into the high-resolution H$\alpha$ contrast effect, i.e., how the ratio of in-transit
to out-of-transit flux as a function of wavelength changes based on which
portion of the star is being occulted by the planet. This is critical to understanding how the surface of an active star such
as HD 189733 will affect the $W_{H\alpha}$ measurement, especially in light of the recent suggestion
by \citet{barnes16} that the H$\alpha$ signal is due mostly to variations on the stellar surface
and not to absorption by circumplanetary material.

\subsection{Model overview}

Here we provide a brief overview of the steps involved in creating the model transmission spectra.
The stellar and planetary parameters used in the model are given in \autoref{tab:starpars}. 
The model has eleven input parameters which are listed in \autoref{tab:modpars}. 
\autoref{tab:modpars} also shows the range of parameter values
we explored with the contrast model but most of the examples presented below focus on a much
smaller range or even specific values. This is due to the fact that some parameters have little influence
on the contrast effect. 

We first simulate the location of spots and faculae on the stellar disk. We
assume that each spot has some concentric associated facular region that has an area equal to 60\%
of the spot area. After all spot locations have been determined, additional facular regions are added to
match the desired spot-to-facular ratio $Q=A_{sp}/A_{fac}$. The location of a spot is determined
by randomly selecting from a normal distribution in latitude centered on the latitude input parameter. The FWHM of the
distribution is kept fixed at 10$^\circ$. The size of the spot is then determined by randomly selecting from a
uniform distribution with boundaries specified by $r_{sp}^{min}$ and $r_{sp}^{max}$, the minimum and maximum
allowed spot radii in units of $R_p$. We do not allow overlap of spots. Finally, the
projected boundaries, which are ellipses, of the spot and surrounding faculae are determined based on the
central spot location. The additional facular regions are added in an identical procedure except the edges
of the ellipses may overlap to allow potentially more irregular patterns.

\begin{deluxetable*}{lccc}
\tablecaption{Stellar and planetary parameters \label{tab:starpars}}
\tablehead{\colhead{Parameter}&\colhead{Symbol}&\colhead{Value$^{a}$}&\colhead{Units}}
\colnumbers
\tabletypesize{\scriptsize}
\startdata
Stellar radius & $R_*$ & 0.756 & $R_\Sun$\\
Stellar rotational velocity & $v$sin$i$ & 3.10 & km s$^{-1}$ \\
Impact parameters & $b$ & 0.680 & $R_*$ \\
Orbital period & $P_{orb}$ & 2.218573 & days \\
Orbital velocity & $v_{orb}$ & 151.96 & km s$^{-1}$  \\
Orbital semi-major axis & $a$ & 0.03099 & AU \\
Planetary radius & $R_p$ & 1.138 & $R_J$ \\
\enddata
\tablenotetext{a}{With the exception of $v$sin$i$, all stellar and planetary parameters taken from \citet{torres08}. The 
$v$sin$i$ value is taken from \citet{collier10}.}
\end{deluxetable*}

\begin{deluxetable*}{lcc}
\tablewidth{1.99\textwidth}
\tablecaption{Contrast model parameters and explored values\label{tab:modpars}}
\tablehead{\colhead{Parameter description}&\colhead{Symbol}&\colhead{Value range}}
\colnumbers
\tabletypesize{\scriptsize}
\startdata
Spot coverage fraction & $A_{sp}$ & 0.005-0.10  \\
Ratio of spots to faculae & $Q$ & 2.0-0.2  \\
Filament coverage fraction & $A_{fil}$ & 0.005-0.03  \\
Ratio of facular to photosphere core H$\alpha$ & $q_{fac}$ & 0.1-6.0  \\
FWHM of facular emission & $\sigma_{fac}$ & 10-50 km s$^{-1}$  \\
Central latitude of active region distribution & $\theta_{act}$ & 0$^\circ$-45$^\circ$ \\
Temperature difference between spots and photosphere & $\Delta T_{sp}$ & 300-800 K \\
Temperature difference between faculae and photosphere & $\Delta T_{fac}$ & 20-50 K \\
Filament H$\alpha$ core contrast & $c_{fil}$ & 0.2-0.8 \\
Minimum spot radius& $r_{sp}^{min}$ & 0.1-0.5 $R_p$ \\
Maximum spot radius & $r_{sp}^{max}$ & 0.3-1.0 $R_p$ \\
\enddata
\end{deluxetable*}

We also include filaments which appear in absorption against the stellar disk \citep[e.g., ][]{heinzel06,kuckein16}. 
The filaments have a fixed width of 0.2 $R_p$ and their length is defined by the total filament coverage
fraction. Filaments are constructed by choosing a random starting point and direction and then letting
the filament grow in that direction with a narrow cone defining the directions each new
piece of the filament is allowed to move towards. The filament is allowed to grow until the desired
coverage fraction is achieved to within 0.1\%. 

\begin{figure}[htbp]
   \centering
   \includegraphics[scale=.56,clip,trim=62mm 26mm 25mm 35mm,angle=0]{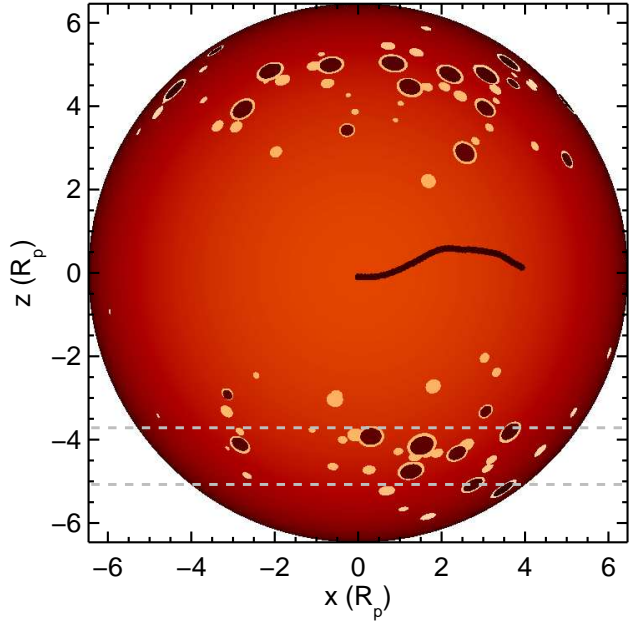} 
   \figcaption{Example stellar disk for a contrast model with $A_{sp}=0.02$, $Q=$0.7, and $A_{fil}=$0.005. 
   The faculae are shown as white annuli surrounding the dark brown spots and randomly placed
   circles with the same latitudinal distribution as the spots. A dark filament can be seen across disk center. 
   The transit path of HD 189733 b is shown by the dashed gray lines. Note that the
   limb-darkening, and limb-brightening in the case of the faculae, is meant to be representative
   of how the features contribute near the H$\alpha$ line core (i.e., absorption or emission) and not to show 
   the precise intensity ratios calculated as a function of wavelength in the models. \label{fig:exgrid}}
\end{figure}

To simulate the transmission spectrum, we first calculate the out-of-transit spectrum by summing the spectra from
individual grid points across the entire stellar surface. The grid resolution is 0.02 $R_p$ which
results in a 647$\times$647 Cartesian grid. The surface feature spectra are PHOENIX model spectra with the addition
of a Gaussian emission profile or extinction of the photospheric spectrum, assuming a Gaussian profile shape,
depending on the type of feature (e.g., a plage or filament). 
The emission and absorption profiles are assumed to be Gaussian due to the approximately
Gaussian shape of the observed H$\alpha$ transmission spectra. In addition, the exact line shape
is less important than the magnitude of the effect, which will not change significantly if a Lorentzian or Voight 
profile is assumed. The spectrum at each point on the stellar disk is shifted by the appropriate stellar rotational velocity, for which we assume rigid rotation.
We also compute the radial velocity of the star relative to the observer \citep[see Eq. 11 of][]{lovis10} and apply
this velocity shift to each of the stellar spectra. We then calculate the stellar spectrum blocked by the planet at 5 minute intervals across the transit. 
The out-of-transit spectrum is subtracted from the blocked spectrum and then the result is divided by the out-of-transit
spectrum to generate synthetic transmission spectra according to \autoref{eq:strans}. An example
of the stellar surface for the parameters $A_{sp}=0.02$, $Q=$0.7, and $A_{fil}=$0.005 is shown in \autoref{fig:exgrid}.

\subsection{Surface feature spectra}

\begin{figure}[htbp]
   \centering
   \includegraphics[scale=.51,clip,trim=42mm 20mm 25mm 35mm,angle=0]{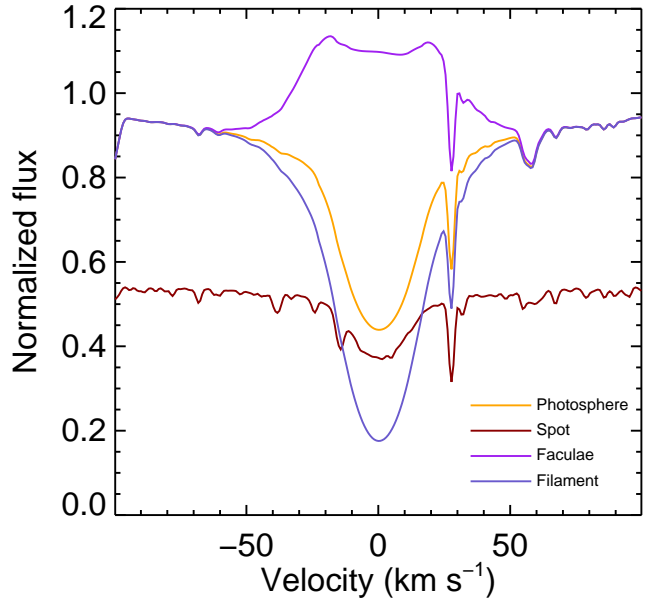} 
   \figcaption{Model surface feature H$\alpha$ spectra at $\mu=1$ for $T_{sp}$=4300 K, $q_{fac}$=1.5, 
   $T_{fac}$=5040 K, and $c_{fil}$=0.6. Each spectrum is normalized by fitting a line 
   to the flux at $-200$ and $+200$ km s$^{-1}$. Note that these are spectra of \textit{individual} surface
   elements and do not represent the absolute fluxes contributing to the explored contrast examples.
   The core emission in the facular spectrum (purple line)
   completely fills in the photospheric absorption in this case. We note that this is more typical of a
   flare rather than a quiescent facular region and is shown here to illustrate the differences between
   the spectra. The temperature difference between
   the spot and the photosphere results in $\sim$45\% less flux from the spot (dark red line). The
   filament (dark blue line) shows strong H$\alpha$ core absorption superimposed on the photospheric
   spectrum. The narrow feature near 30 km s$^{-1}$ is a \ion{Cr}{1} line. We ignore the absorption
   contribution from this line when calculating $W_{H\alpha}$. \label{fig:exspecs}}
\end{figure}

Each distinct surface feature contributes a different H$\alpha$ spectrum. We include four separate
surface components: 1. the naked photosphere; 2. star spots; 3. faculae; and 4. filaments.
Below we discuss each spectrum in detail, as well as the other parameters associated with that specific
surface feature. Example H$\alpha$ spectra of the four surface components at $\mu=\cos{\theta}=1.0$,
where $\theta$ is the angle between the normal vector to the stellar surface and the line-of-sight,
are shown in \autoref{fig:exspecs} for a model with
$T_{sp}$=4300 K, $q_{fac}$=1.5, $T_{fac}$=5040 K, and $c_{fil}$=0.6. The faculae are approximately
the same brightness as the photosphere at $\mu$=1.0 (see \autoref{eq:lbfac}). 
The normalized continuum flux for the spot spectrum is reduced by a factor of $(T_{sp}/T_{eff})^4$.
We ignore all small-scale velocity effects in the spectra, such as photospheric convective blue shifts,
since we are only concerned with the overall contrast between the different regions. The contrast is
weakly affected by these velocity shifts, which are of the order $\sim$200-300 m s$^{-1}$ \citep[e.g.,][]{meunier10b,lanza11}.

\subsubsection{Photospheric spectra}

For the photosphere we use a PHOENIX model spectrum \citep{husser13} with $T_{eff}$=5000 K, log$g$=4.5,
and [Fe/H]=0.0. It is important to emphasize that the choice of photospheric spectrum is not critical to exploring
the contrast effect since most of the stellar surface is dominated by the photosphere. Furthermore, the spotted,
filament, and facular spectra are scaled by or built from the photospheric spectrum in some manner so the entire
surface is constructed relative to the photosphere. Thus a choice of the model $T_{eff}$ of 4500 K or 5500 K 
does not significantly affect our conclusions. We note that \citet{boyajian15} recently measured $T_{eff}$ for
HD 189733 to be 125 K lower (4875$\pm$43 K) than the value used here.  

\subsubsection{Spot spectra}

For the star spot spectra we use PHOENIX model spectra with the same log$g$ and [Fe/H] values but
with varying temperatures in step sizes of 100 K. Spots are
cooler than the surrounding photosphere with differences in temperature that depend on spectral
type, although this dependence may break down at $T_{eff}>T_\odot$ \citep{eker03,berdyugina05}.  We
note that we do not distinguish between umbral and penumbral temperatures; they are averaged into a
single spot temperature. \citet{herrero16} demonstrate (their Figure 2) the good agreement between
observed solar spot spectra and model PHOENIX spectra at a cooler $T_{eff}$.
 
\citet{pont13} provided a detailed analysis of spot crossing events during HD 189733 b transits. Their main
conclusions were: 1. there is no evidence for the transiting of concentrated bright faculae, suggesting
that facular regions are spread out across the stellar surface; 2. the spot filling factor is $\sim$1-2\%; and 3. 
the temperature difference between the photosphere and spots is $\Delta$$T=750$ K. Photometric modeling
performed by \citet{herrero16} suggests that spots dominate changes in the stellar brightness across stellar
rotation periods. The observations presented in \citet{pont13} span $\sim$6 years which provides evidence 
that the spot filling factor does not change significantly on long time scales. Their results provide reasonable 
spot parameter values around which to base our investigation.

\subsubsection{Filament spectra}

Filament spectra are identical to photospheric spectra but
with the H$\alpha$ line core subject to further absorption. \citet{kuckein16} measure the contrast across
H$\alpha$ from $\sim$$-80$ to $+80$ km s$^{-1}$ for filaments on the solar disk. They find $\sim$60\% 
less flux in the line core compared with the bare photosphere, although the values vary depending on the where the measurement is
made on the filament. We adopt a single contrast value of 0.6 (i.e., 60\% less flux) at the line core for the filament spectra. The absorption
is modeled as a Gaussian with a full width at half maximum (FWHM) of 40 km s$^{-1}$. We do not
consider prominences, which are filaments projected beyond the edge of the stellar disk, although
large prominences may produce pre- or post-transit contrast signals.

\subsubsection{Facular spectra}

Facular spectra consist of the intensity-weighted sum of the underlying
photosphere and overlying emission region. During small solar flares, the ratio of H$\alpha$ line core
emission to the underlying photospheric spectrum can reach $\sim$2.0-4.0 \citep{johnskrull97,xu16}. The ratios
measured for non-flaring faculae are closer to $\sim$1.2 \citep{ahn14} and very weak flares
show ratios of $\sim$1.1-1.6, or $q_{fac}=0.1-0.6$ \citep{deng13}. For the contrast models, we are only interested in the
low-level stellar activity that does not change dramatically over the course of a single night. This implies
values for $q_{fac} \sim 0.1-2.0$. Larger ratios than this likely do not
apply to non-flaring regions. However, we explore larger values of $q_{fac}$ since larger values
are needed to reproduce the observed $W_{H\alpha}$ values. Whether or not these
large values for the H$\alpha$ core emission are physical will be discussed below.

Facular and plage regions are limb-brightened relative to the underlying photosphere.
We adopt the limb-brightening law of \citet{meunier10a} and \citet{herrero16}:

\begin{equation}\label{eq:lbfac}
  c_{fac}(\mu) = \left( \frac{T_{eff} + \Delta T(\mu)}{T_{eff} + \Delta T_{fac}} \right)^4
\end{equation}

\noindent where $\Delta T_{fac}$ is the temperature difference between the photosphere and faculae
and $\Delta T(\mu)=250.9-407.4\mu+190.9\mu^2$.

\subsection{Contrast and limb-darkening/brightening}
 
\citet{czesla15} calculated differential center-to-limb variations (CLV), or differences in the limb-darkening
or brightening as a function of wavelength across a specific spectral line, for the \ion{Ca}{2} H and K and
\ion{Na}{1} D lines for HD 189733. They demonstrated that the \ion{Na}{1} lines show limb-\textit{brightening} in the
wings of the line compared with the line core or neighboring continuum. This is important since a transiting
planet with no atmosphere can produce a transmission signal, even if there are no active regions present
on the stellar surface, due to the fact that the ratio of the line core to
the continuum changes as a function of the transit \citep[see Figures 1-5 of][]{czesla15}. This effect was
recently included in \ion{Na}{1} transit modeling by \citet{khalafinejad}.

Such CLVs will also affect the in-transit H$\alpha$ measurements. In order to account for this
effect, we have calculated high-resolution H$\alpha$ spectra using the program
SPECTRUM\footnote{http://www.appstate.edu/~grayro/spectrum/spectrum.html} by \citet{gray94} and an
ATLAS9 model atmosphere\footnote{http://wwwuser.oats.inaf.it/castelli/} with $T_{eff}$=5000 K,
log$g$=4.5, and [Fe/H]=0.0. The spectra were computed at fourteen values of $\mu=cos(\theta)$
between 0.01 and 1.0. To avoid interpolation during the contrast model calculations, we fit the
following limb-darkening law from \citet{hestroffer98} to each wavelength across the spectrum:

\begin{equation}\label{eq:limbdark}
  I(\mu)=1-u(1-\mu^\alpha)
\end{equation}

\begin{figure}[htbp]
   \centering
   \includegraphics[scale=.41,clip,trim=20mm 10mm 30mm 25mm,angle=0]{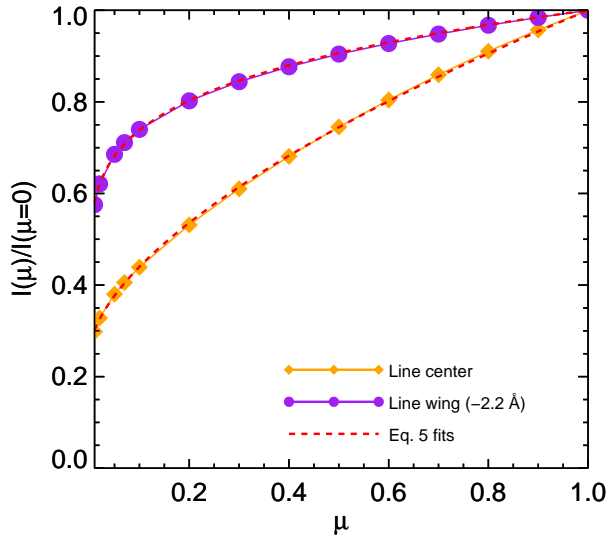} 
   \figcaption{Examples of wavelength-dependent limb-darkening curves from the synthetic spectra (solid
   lines and symbols) and the fits using \autoref{eq:limbdark} (red dashed lines). Note the steeper limb-darkening
   in the line core compared with the line wing. \label{fig:ldcurves}}
\end{figure}

\noindent The parameters from \autoref{eq:limbdark} are then used to compute $I(\mu)$ as a function of wavelength
across the line for a densely sampled grid of $\mu$ across the stellar disk. Examples of the
$I(\mu)$/$I(\mu=1)$ vs. $\mu$ curves and the corresponding \autoref{eq:limbdark} fits are shown in \autoref{fig:ldcurves}.
There is a significant difference between the limb-darkening in the core of the line versus the line wing which
can impact the measured $W_{H\alpha}$ values.

\begin{figure*}[htbp]
   \centering
   \includegraphics[scale=.75,clip,trim=12mm 55mm 20mm 65mm,angle=0]{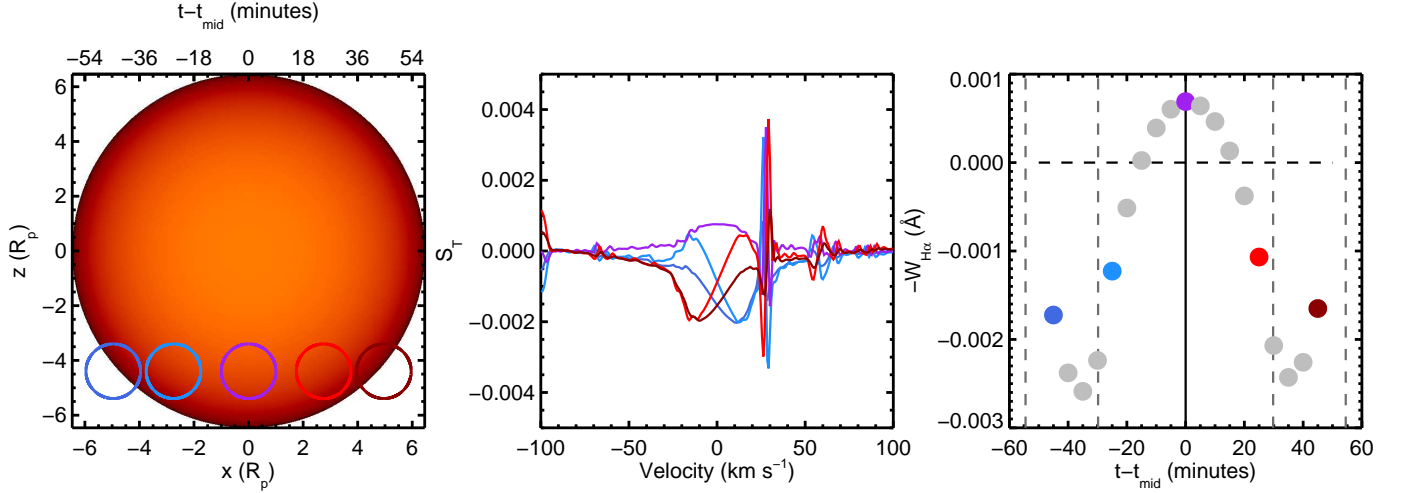} 
   \figcaption{Example of the contrast effect for a star with no spots, faculae, or filaments. The planet's
   position at five in-transit times is shown with the solid circles (left panel). The top x-axis gives the time from mid-transit
   corresponding to the distances on the bottom x-axis. The transmission spectra $S_T$ for each in-transit time are
   shown in the middle panel. The narrow spikes near 27 km s$^{-1}$ are the result of a \ion{Cr}{1} line at 6563.40 \AA. 
   Note that this line is ignored in the calculation of $W_{H\alpha}$.
   $S_T$ shows a shift from stronger to weaker absorption as the planet approaches mid-transit. This is the result
   of CLVs where the difference between the limb-darkening in the wing versus the line core is smallest near the
   middle of the stellar disk. Note the difference in scale between the right panel and the transmission spectra
   shown in \autoref{fig:hatransits}. The CLV effects are an order of magnitude lower than the measured
   signals. The right panel shows the $W_{H\alpha}$ values calculated for the full set of in-transit times. 
   The vertical dashed lines mark the transit contact points and the vertical solid line marks mid-transit.
   The strongest absorption signal, or in this case contrast signal, is when the planet
   is near the edge of the stellar disk but still almost completely covers the limb. 
   This is very similar behavior to the \ion{Na}{1} D calculations presented
   by \citet{czesla15}. \label{fig:nospots}}
\end{figure*}

\autoref{fig:nospots} shows a transit example for a pure photosphere, i.e., no spots, faculae, or filaments.
The only mechanism affecting the transmission spectrum, or, in the absence of a planetary atmosphere, the contrast
spectrum, is the CLVs described above. Five representative in-transit times are shown in the left panel of \autoref{fig:nospots}
and their corresponding $S_T$ profiles are shown in the middle panel. The deepest contrast profiles are seen when the
planet is near the stellar limb, but still almost completely covers the disk, since this is where the limb-darkening curves in the line wing and line core differ the
most. The $W_{H\alpha}$ values for the entire in-transit calculation are shown in the right panel of \autoref{fig:nospots} which
explicitly shows the ``absorption'' effect near the stellar limb. 

An important takeaway from the example shown in \autoref{fig:nospots} 
is that the signal induced by CLVs at H$\alpha$ is well below the measured
$W_{H\alpha}$ values for most of the transit presented here. This is especially true for values of $W_{H\alpha}$ 
near mid-transit where the CLV effect is weakest. Although HD 189733 is an active star and thus it is unlikely
that the visible hemisphere is ever pure photosphere, this baseline demonstrates
that the observed H$\alpha$ transit signals cannot be caused by only CLVs in this specific case. However,
the magnitude of the CLV effect is significant upon ingress and egress and must be included in any model
of the in-transit absorption. It is also important to
highlight the velocities of the $S_T$ profiles in the middle panel of \autoref{fig:nospots}: upon ingress, $S_T$ 
shows red-shifted velocities near maximum absorption while upon egress it shows blue-shifted velocities. This
contrast signal will have an important effect on the $v_{H\alpha}$ models presented in \autoref{sec:velocities}. 

\subsection{A note on parameter importance}

Specifying unique spectra and a unique spatial configuration for active regions on the stellar disk 
requires all of the parameters in \autoref{tab:modpars}. However, some of 
these parameters are much more important than others in determining changes in the contrast spectra. In addition,
we can look to the observed H$\alpha$ transmission spectra and previous HD 189733 studies 
for guidance on other parameter values. For example, the spot radii actually govern the 
distribution of spots since larger spot radii means fewer spots for
a constant $A_{sp}$. But $A_{sp}$ is more important for determining the relative weights between spotted
spectra and the photosphere, which in turn is more important for the contrast spectrum. 
The distribution of active regions is a strong determinant of the time series but does not
strongly affect the contrast spectrum if the planet is not occulting a large fraction of the active regions.
Another example is $\sigma_{fac}$ of the active region emission. 
With the assumption of the spectra being produced by the contrast effect, we can infer that
$\sigma_{fac}\approx40$ km s$^{-1}$. This suggests that we do not need to explore a large range of values 
for $\sigma_{fac}$ since most values will not be able to reproduce the line profile shape. 
For these reasons, although we have investigated the full range of parameters given in \autoref{tab:modpars},
we do not present details of the full extent these efforts and instead focus on illuminating cases that
are most relevant to the measurements.    

\subsection{Active stellar surface transits}

In this section we present various illustrative examples of the contrast effect for an active stellar surface.
These examples are not meant to be exhaustive but rather representative of how general
configurations affect the contrast spectrum and which parameters are most important in
producing a significant contrast spectrum. We do not explore scenarios where the active regions
are isolated to a small portion of the stellar disk since transits of these isolated regions have
short durations. Instead, we explore scenarios that might produce absorption across the entire transit.

\subsubsection{Transiting a photospheric chord}

\begin{figure*}[htbp]
   \centering
   \includegraphics[scale=.75,clip,trim=12mm 55mm 20mm 65mm,angle=0]{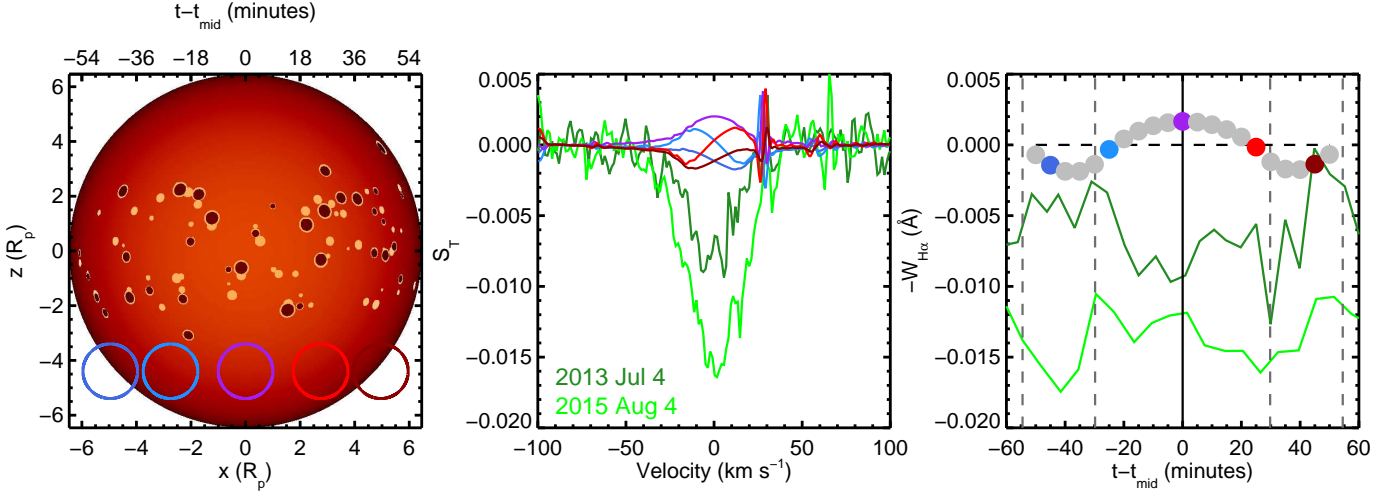} 
   \figcaption{Transit of a photospheric chord with $A_{sp}=0.02$, $Q=0.7$, $A_{fil}=0.0$,
   $\theta_{act}=\pm 10^\circ$, and $q_{fac}=1.5$. The 2013 Jul 4 and 2015 Aug 4 Keck data are
   shown in green in both the middle and right panels. Note the different scale compared with
   \autoref{fig:nospots}. Since the in-transit spectrum is weighted towards the active
   regions, the $W_{H\alpha}$ curve (right panel) is shifted up due to the line core being filled in by facular 
   emission. This can also be seen in $S_T$ (middle panel) where the spectra show stronger emission
   features. \label{fig:phottran}}
\end{figure*}

The planet can either transit active regions or the photosphere. Either way, the in-transit spectrum will be
altered relative to the out-of-transit spectrum. \autoref{fig:phottran} shows an example of the
planet transiting a chord with no active regions for the parameter values $A_{sp}=0.02$, $Q=0.7$, $A_{fil}=0.0$,
$\theta_{act}=10^\circ$, and $q_{fac}=1.5$. The CLVs can still be seen in the contrast profiles (middle panel)
but since the in-transit spectra are weighted towards active regions, $S_T$ shows emission in the
line core at most $t-t_{mid}$ rather than absorption. This also shifts the $-W_{H\alpha}$ curve up towards
less absorption. 

\begin{figure*}[htbp]
   \centering
   \includegraphics[scale=.75,clip,trim=12mm 55mm 20mm 65mm,angle=0]{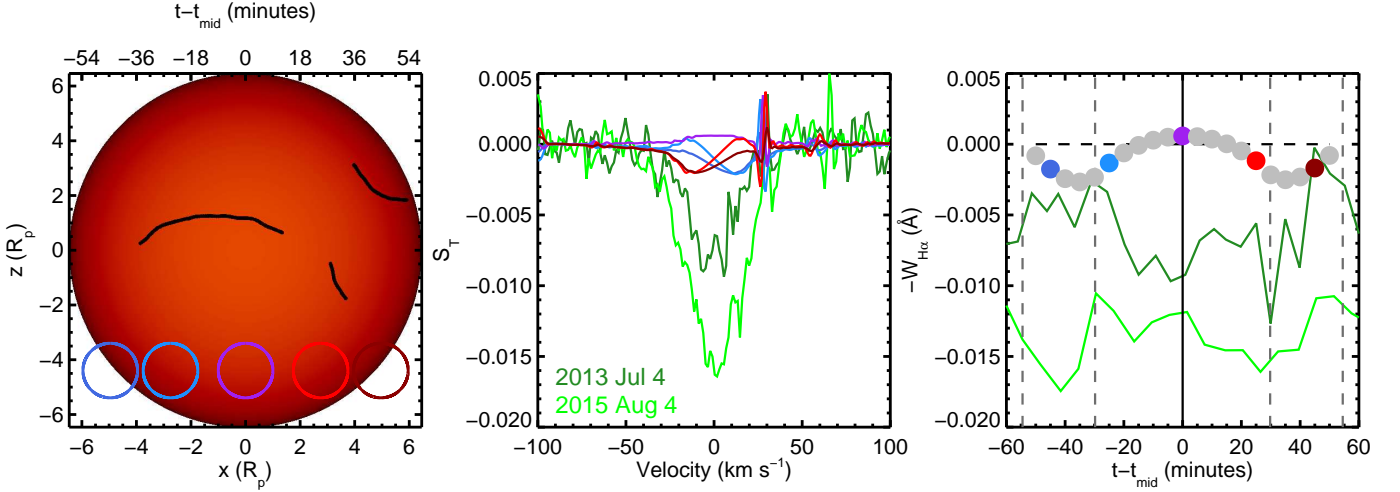} 
   \figcaption{Same as \autoref{fig:phottran} but for the pure filament case with $A_{fil}=0.01$. The filaments produce a very weak contrast effect. 
   \label{fig:purefil}}
\end{figure*}

One important thing to reiterate in this example is that strong absorption lines are not produced
in the contrast spectra. This occurs for two reasons: 1. the facular core emission dominates 
the contrast spectrum; 2. the spot spectrum actually has a \textit{smaller core-to-continuum contrast
compared with the photospheric spectrum}. This reflects the fact that the H$\alpha$ line strength
\textit{decreases} for cooler stellar spectra. In this example, $\Delta T_{sp} = 700$ K so the spot
spectrum is identical to the spot spectrum in \autoref{fig:exspecs}.
This has important consequences for the pure-spot scenario, which we do not show: even without the core emission
from the facular/plage regions, the contrast spectrum would still show core emission instead
of absorption since the spot spectrum shows less core absorption than the photosphere. 
As a result, spots have little effect on the H$\alpha$ transmission spectrum.

We also show the filament-only case in \autoref{fig:purefil} for $A_{fil}=0.01$. Filaments
do not produce strong contrast spectra for reasonable values of $A_{fil}$, although we note that the average transmission
spectrum (not shown) for an entire transit is seen in weak absorption for the case of the photospheric chord transit.
Individual transmission spectra are seen in emission for transits of filaments. 
Since their contribution to the contrast effect is minor, even when the planet transits filaments,
we do not focus on them further.

\subsubsection{Transiting an active latitude}

The photospheric transit examples demonstrate that $W_{H\alpha}$ cannot
be measured in absorption if the planet does not consistently occult facular regions. Thus in
order to produce the strongest $W_{H\alpha}$ curves shown in \autoref{fig:hatransits} HD 189733 b
must be transiting a chord that is densely covered with faculae and plage regions. Furthermore,
as we demonstrate below, the emission strength in these facular regions
must be similar to what is seen in flaring regions on the Sun.

\begin{figure*}[htbp]
   \centering
   \includegraphics[scale=.75,clip,trim=12mm 50mm 20mm 65mm,angle=0]{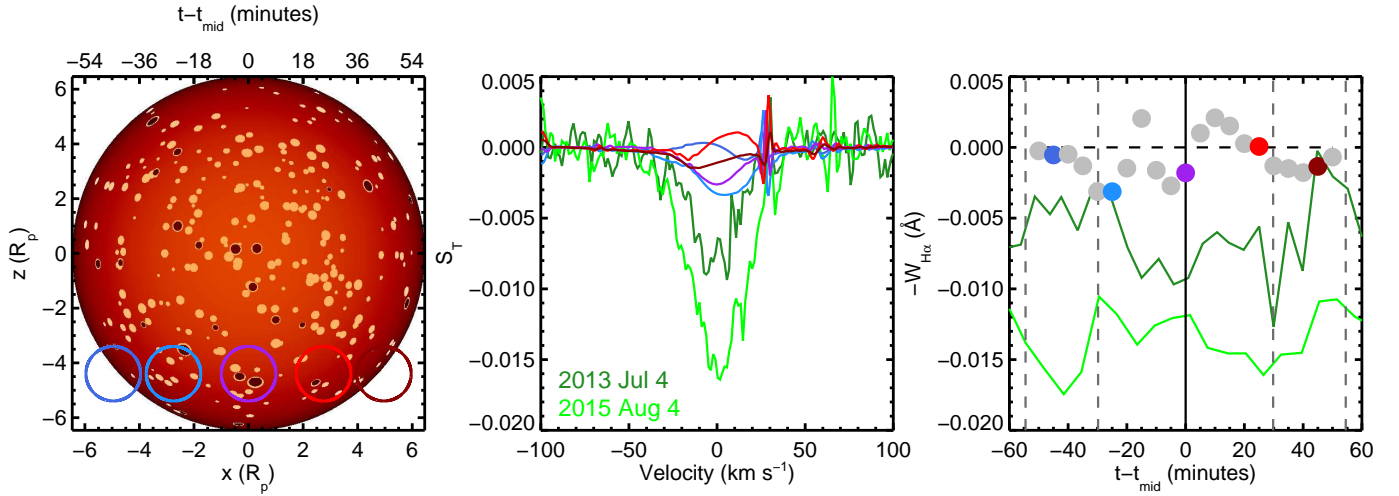} 
   \figcaption{Same as \autoref{fig:phottran} but now with $A_{sp}=0.01$, $Q=0.13$, $A_{fil}=0.0$,
    and $q_{fac}=4.0$ for a uniform distribution of spots and facular regions. $W_{H\alpha}$ (right panel)
    is erratic since the planet never occults a large area of active regions. Note that $q_{fac}$ is very
    large in this example but very similar behavior is seen for smaller values, although the variations
    in $W_{H\alpha}$ are smaller. \label{fig:actuni}}
\end{figure*}

Here we present examples of the planet transiting chords that contain active regions. The first
scenario is shown in \autoref{fig:actuni} for uniformly distributed active regions with $A_{sp}=0.01$,
$Q=0.13$, $A_{fil}=0.0$, and $q_{fac}=4.0$. The $W_{H\alpha}$ timeseries in the right panel
is erratic; there is no trend. We have chosen the large value of $q_{fac}=4.0$ to demonstrate
that even very strong core emission cannot produce strong contrast spectra (center panel) if
the planet never occults significant areas of active regions.

\begin{figure*}[htbp]
   \centering
   \includegraphics[scale=.75,clip,trim=12mm 55mm 20mm 65mm,angle=0]{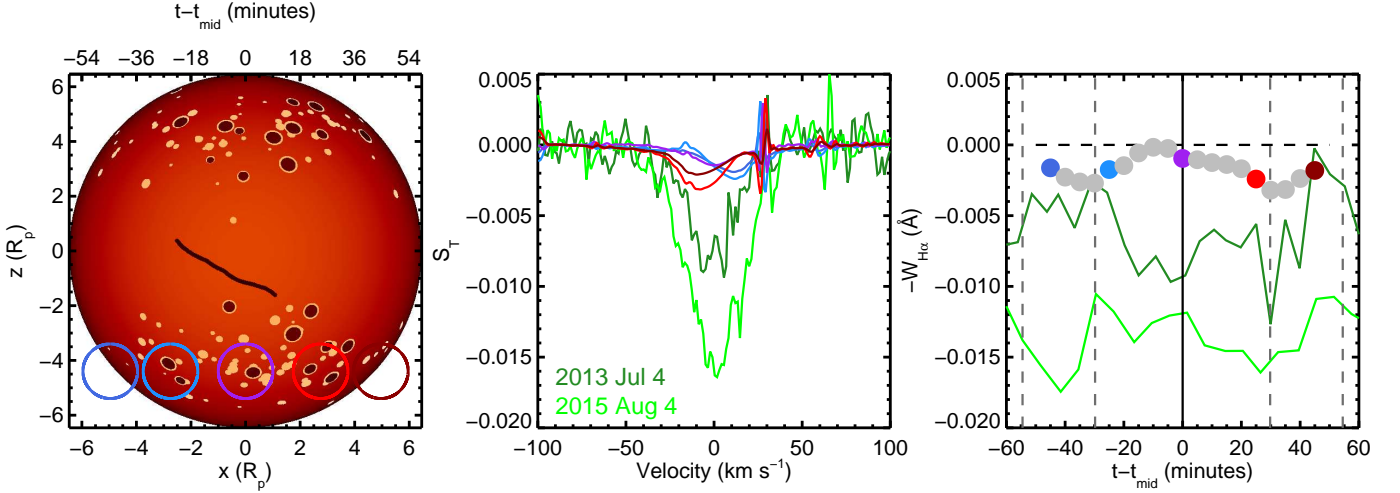} 
   \figcaption{Same as \autoref{fig:phottran} but now with $A_{sp}=0.02$, $Q=0.5$, $A_{fil}=0.005$,
   $\theta_{act}=\pm 40^\circ$, and $q_{fac}=0.5$. Now the in-transit spectrum is weighted towards
   the photosphere since the planet occults active regions almost continuously throughout the transit.
   The contrast is strongest when the highest active region area is occulted, e.g., during the purple
   and red planet positions in the first panel. The $W_{H\alpha}$ values show absorption,
   although there is an abrupt shift near mid-transit when the planet begins to occult more of the
   facular regions. \label{fig:actlow}}
\end{figure*}

\begin{figure*}[htbp]
   \centering
   \includegraphics[scale=.75,clip,trim=12mm 55mm 20mm 65mm,angle=0]{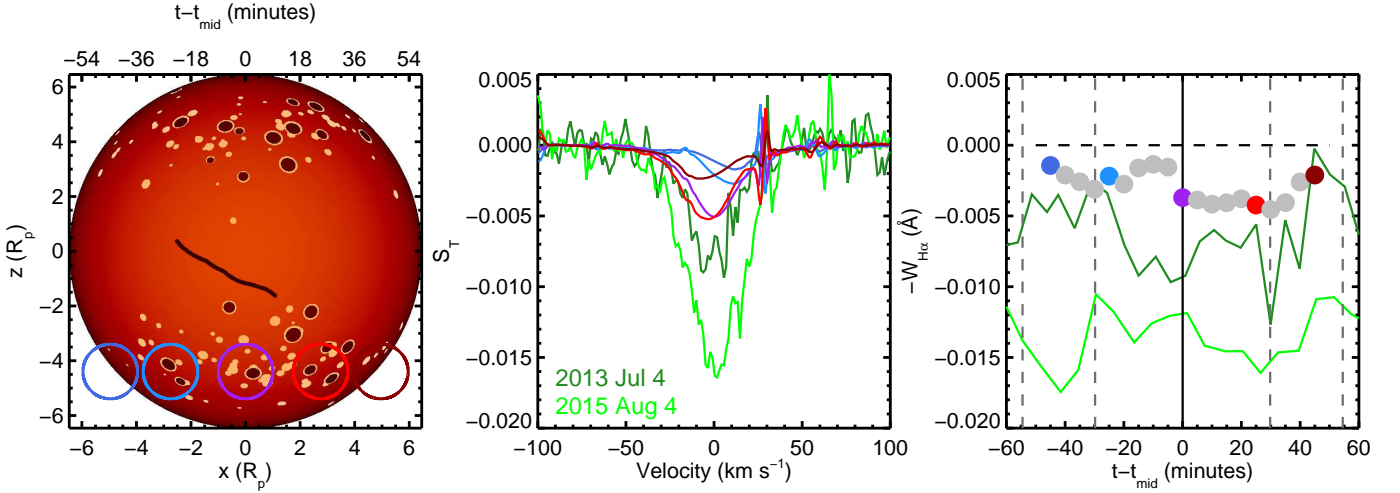} 
   \figcaption{Same as \autoref{fig:actlow} but now with $q_{fac}=1.5$. The stellar surface in the first panel is identical to that in
   \autoref{fig:actlow}. The $W_{H\alpha}$ values show continuous absorption,
   although there is an abrupt shift near mid-transit when the planet begins to occult more of the
   facular regions. Although the contrast spectrum shows absorption throughout the transit,
   $W_{H\alpha}$ is still 2-3$\times$ smaller than is measured in most of the full transits from
   \autoref{fig:hatransits}. \label{fig:acthigh}}
\end{figure*}

\autoref{fig:actlow} and \autoref{fig:acthigh} show examples of the planet transiting an active latitude.
The active region configuration is identical in both cases. The only difference is in the value of $q_{fac}$:
in \autoref{fig:actlow} $q_{fac}=0.5$, while in \autoref{fig:acthigh} $q_{fac}=1.5$. The mid-transit times
where the planet occults the largest area of facular regions show significant differences in the
strength of the transmission spectrum: for the larger $q_{fac}$ the absorption line is $\sim$2$\times$
stronger. Furthermore, the $W_{H\alpha}$ values consistently show absorption across the entire
transit whereas in the $q_{fac}=0.5$ case $W_{H\alpha}$ approaches zero near mid-transit.
These examples illustrate the important point that \textit{the strength of $q_{fac}$ is the dominant
factor in determining the magnitude of the contrast effect for any given transit snapshot.} For the
$W_{H\alpha}$ timeseries, $Q$, and therefore the facular coverage fraction, is also important.
But $q_{fac}$ is the main determinant of the depth of the contrast spectrum. In neither case does
the contrast effect produce absorption similar to what is observed.

\begin{figure*}[htbp]
   \centering
   \includegraphics[scale=.7,clip,trim=12mm 55mm 20mm 70mm,angle=0]{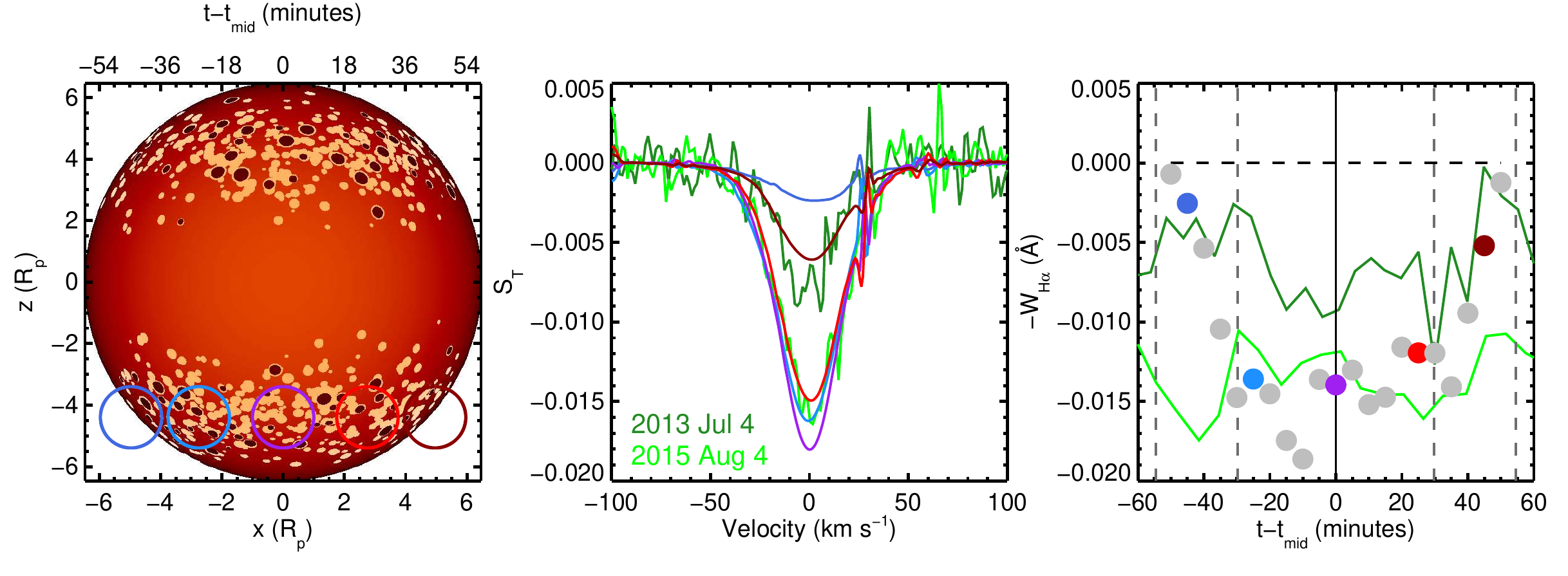} 
   \figcaption{Same as \autoref{fig:nospots} but now with $A_{sp}=0.03$, $Q=0.15$, $A_{fil}=0.0$,
   $\theta_{act}=40^\circ$, and $q_{fac}=4.0$. In this case facular/plage regions cover 20\% of the 
   stellar disk. The $W_{H\alpha}$ values show continuous absorption
   that is of comparable strength to the full $W_{H\alpha}$ transits seen in \autoref{fig:hatransits}.
    \label{fig:actvery}}
\end{figure*}

\begin{figure*}[htbp]
   \centering
   \includegraphics[scale=.7,clip,trim=12mm 55mm 20mm 70mm,angle=0]{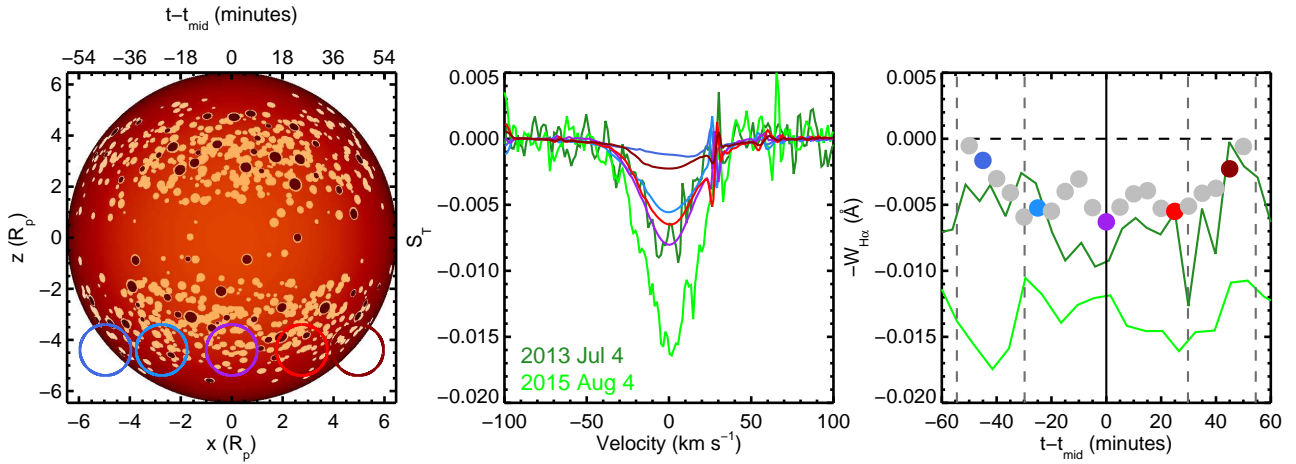} 
   \figcaption{Same as \autoref{fig:actvery} but with $\theta_{act}=30^\circ$, and $q_{fac}=4.0$. 
   The $W_{H\alpha}$ values are much weaker since the planet now transits the edge of the
   active region latitudinal belt. The $W_{H\alpha}$ values in the right panel are shown on the same
   scale as \autoref{fig:actvery} to emphasize the much weaker contrast effect. \label{fig:actvery_30}}
\end{figure*}

The previous two examples demonstrate the need for greater contrast at H$\alpha$
in order to reproduce the observations. \autoref{fig:actvery} and \autoref{fig:actvery_30} show scenarios with very high surface coverage
fractions ($A_{sp}=0.03$ and $Q=0.15$ in both cases) and very strong facular/plage emission
($q_{fac}=4.0$). The only difference is that the active region distribution in \autoref{fig:actvery}
is centered at $\theta_{act}=\pm 40^\circ$ while in \autoref{fig:actvery_30} it is centered at
$\theta_{act}=\pm 30^\circ$. In \autoref{fig:actvery} the planet transits an almost constant
area of active regions and the strength of the facular emission produces strong contrast
absorption lines (center panel) and a relatively uniform $W_{H\alpha}$ timeseries (right panel).
Although $q_{fac}$ is the same in \autoref{fig:actvery_30} the contrast profiles are weaker
and $W_{H\alpha}$ is $\sim3\times$ shallower. This is the result of the planet transiting the
edge of the active region distribution where, at any in-transit snapshot, the planet occults
a relatively smaller area of facular/plage regions compared to the distribution in \autoref{fig:actvery}.
\textit{This suggests that the active regions on HD 189733's surface must be centered very near
the planet's transit chord in order to produce a relatively uniform $W_{H\alpha}$ timeseries.}

\begin{figure*}[htbp]
   \centering
   \includegraphics[scale=.7,clip,trim=12mm 55mm 20mm 70mm,angle=0]{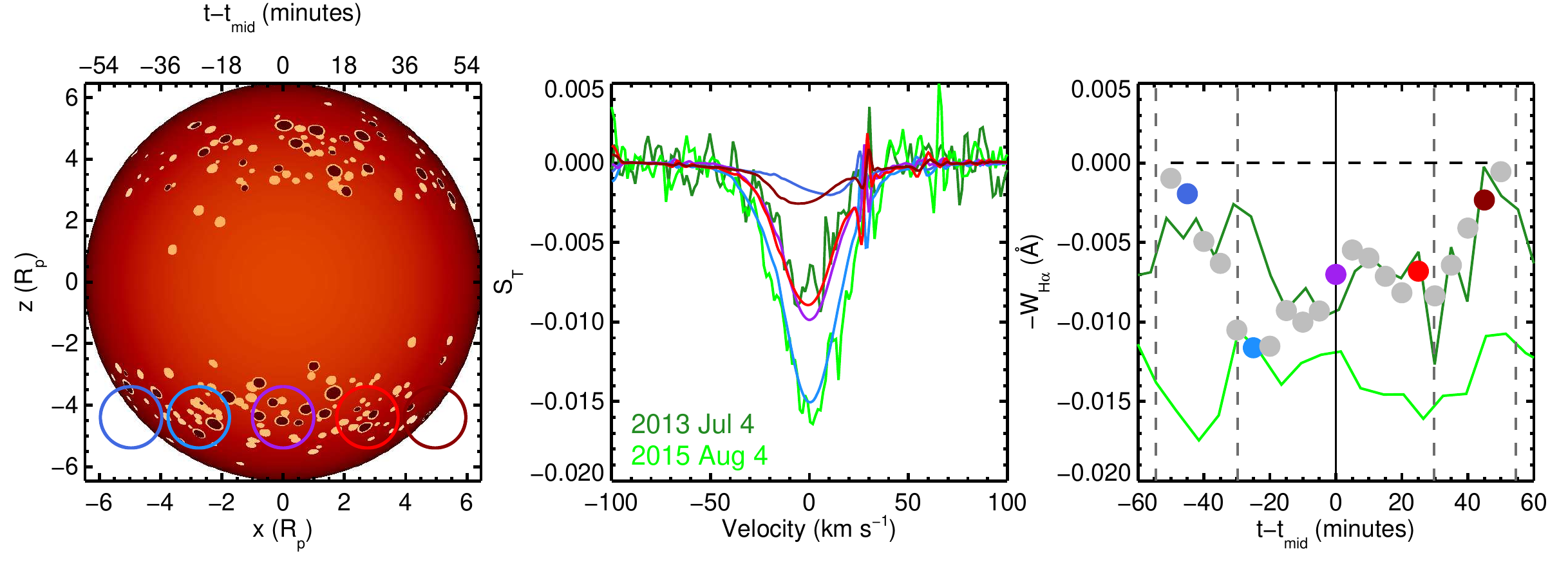} 
   \figcaption{The less active case of $A_{sp}=0.02$, $Q=0.4$, $A_{fil}=0.0$, and $q_{fac}=4.0$. Although
   the facular/plage coverage is much lower than in \autoref{fig:actvery} $W_{H\alpha}$ is consistently
   observed in absorption across the whole transit. \label{fig:actless}}
\end{figure*}

The value of $Q$ in the above examples, which equates to a 20\% surface coverage of facular/plage
regions, does not necessarily need to be so low, i.e., the active region surface coverage does not
need to be so large. \autoref{fig:actless} shows the case for $A_{sp}=0.02$, $Q=0.4$, $A_{fil}=0.0$,
and $q_{fac}=4.0$. While $W_{H\alpha}$ is not as uniform, the values during the first half of the
transit approach the largest observed values from \autoref{fig:hatransits}. We note that moving the
center of the active region distribution to $\pm 30^\circ$ latitude results in essentially no
contrast for the entire transit. This again illustrates the need for the active regions to be
concentrated near the transit chord and to be uniformly distributed in longitude. It also shows that
if $Q$ increases, $q_{fac}$ must remain high or even increase in order to get close to the observed
values; decreasing $q_{fac}$ from 4.0 to $\sim$2.0 produces a very weak contrast effect.

\subsection{Comparisons with low-activity templates}
\label{sec:concomp}

We demonstrated above that it is possible under some conditions to reproduce even the strongest 
observed $W_{H\alpha}$ in-transit signals using the contrast effect. However, the only parameter configurations that are
able to produce these $W_{H\alpha}$ values involve modest to large
facular coverage fractions and $q_{fac} \sim 4-5$. Although we do not know exactly how
the active regions are distributed on HD 189733 or the strength of the facular/plage H$\alpha$
core emission, we can attempt to roughly constrain the \textit{combination} of these parameters
by comparing HD 189733 with less active main sequence templates of similar $T_{eff}$.

We have downloaded archived Keck HIRES data for the three template stars listed in
\autoref{tab:tempstars}. HD 189733 is also listed for comparison. We use these templates to find
parameter combinations of $q_{fac}$ and $A_{fac}$, where $A_{fac}$ is the fractional surface area covered by
facular regions, that can fill in the H$\alpha$ absorption of the less-active template and match the
line profile of HD 189733. Note that $A_{fac}$ is just another way of specifying $Q$ in the contrast
models, where $Q=A_{sp}/A_{fac}$. The HD 189733 spectrum is an average of the comparison spectra used in all of the Keck
transits (purple circles in \autoref{fig:hatransits}).  The facular spectrum $S_{fac}$ is
constructed in the same manner as the contrast model facular spectrum except now $q_{fac}$ refers to
the core flux of the template spectrum. The final spectrum is the weighted average of the
photospheric spectrum $S_{phot}$, which is the observed template spectrum, and $S_{fac}$:

\begin{equation}\label{eq:hacon}
S_{tot} = (1-A_{fac})S_{phot} + A_{fac} S_{fac}
\end{equation}

\begin{deluxetable*}{lcccc}
\tablecaption{Less-active comparison stars \label{tab:tempstars}}
\tablehead{\colhead{}&\colhead{$T_{eff}$}&\colhead{$M_*$}&\colhead{$R_*$}&\colhead{}\\
\colhead{ID}&\colhead{(K)}&\colhead{($M_\odot$)}&\colhead{($R_\odot$)}&\colhead{$S_{HK}$}}
\colnumbers
\tabletypesize{\scriptsize}
\startdata
HD 189733 & 5040 & 0.81 & 0.76 & 0.51 \\
 HD 192263 & 4975 & 0.83 & 0.75 & 0.47 \\
 HD 104067 & 4956 & 0.91 & 0.75 & 0.34 \\
  HD 87883 & 4958 & 0.78 & 0.77 & 0.28 \\
\enddata
\tablecomments{HD189733 parameters taken from \citet{torres08}. All template stellar parameters taken from \citet{valenti05}.
Values of $S_{HK}$ are taken from \citet{isaacson10}.}
\end{deluxetable*}

While all of the observed template spectra are rotationally broadened according 
to the $v$sin$i$ value of HD 189733, we do not account for the CLVs discussed
previously. Thus we are only exploring first-order approximations for which $q_{fac}$
and $A_{fac}$ values are needed to reproduce the HD 189733 H$\alpha$ core. 
Since we do not have spatially resolved spectra of these stars we cannot build
up the spectrum across the stellar disk as is done in the contrast model case.

\begin{figure*}[htbp]
   \centering
   \includegraphics[scale=.75,clip,trim=10mm 10mm 5mm 15mm,angle=0]{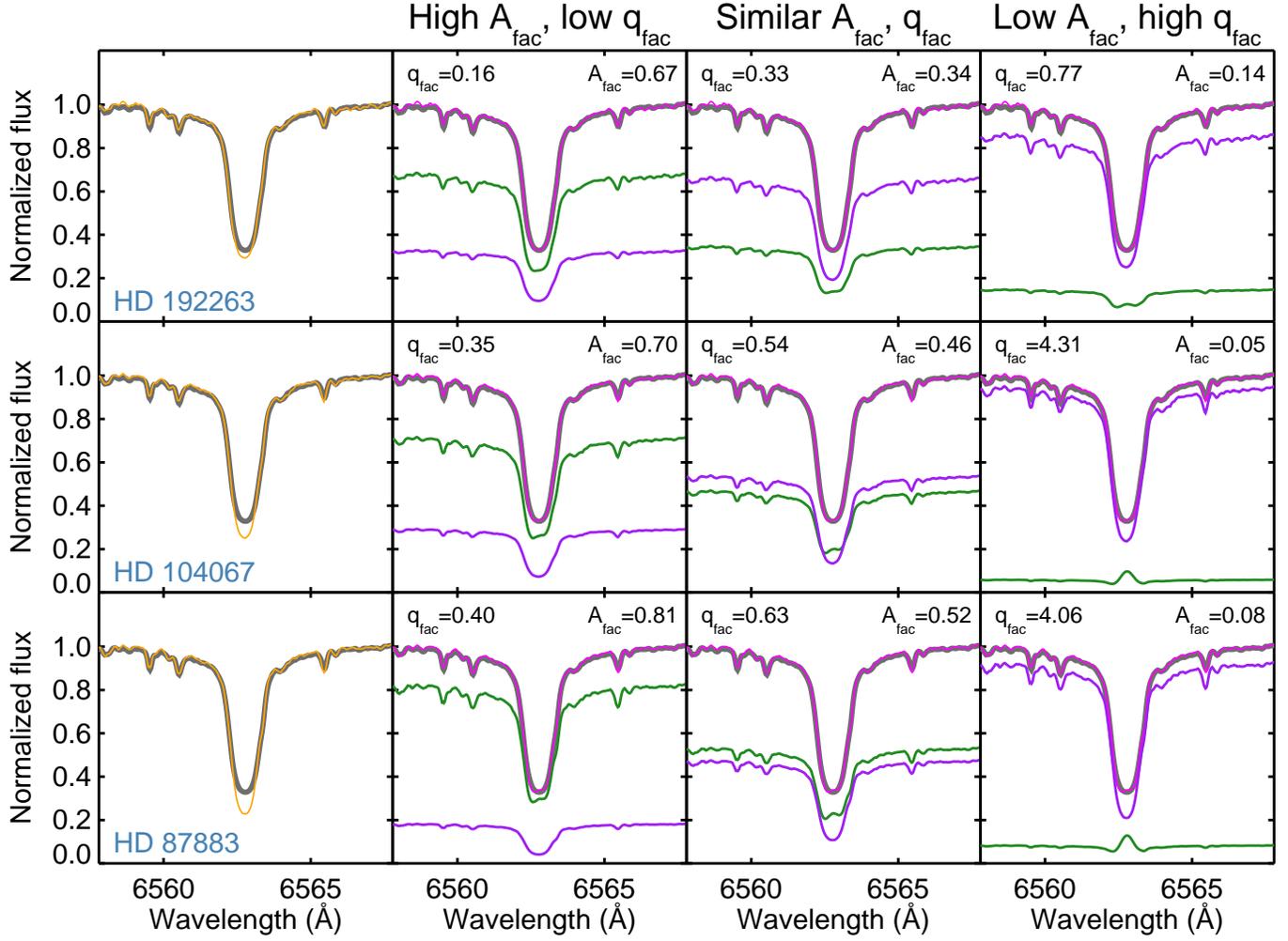} 
   \figcaption{Comparison of HD 189733 H$\alpha$ profile with
   less active main-sequence templates. Column 1 shows the observed template over-plotted in orange on top
   of the HD 189733 spectrum (dark gray line). Columns 2-4 show
   various combinations of $A_{fac}$ and $q_{fac}$ from \autoref{eq:hacon} that are needed to reproduce the
   HD 189733 H$\alpha$ core depth. The photospheric spectrum is shown in purple and the facular/plage
   spectrum is shown in green. A clear relationship is seen for $A_{fac}$ and $q_{fac}$: as $q_{fac}$ increases,
   $A_{fac}$ must decrease to produce the same disk-integrated spectrum. This implies that we cannot use both
   high $A_{fac}$ and high $q_{fac}$ to describe the observed H$\alpha$ transits since it would violate the
   observed disk-integrated core strength. \label{fig:hacorefill}}
\end{figure*}

Three different parameter combinations are shown for each template in \autoref{fig:hacorefill}. 
A $\chi^2$ minimization routine is used to produce the fits for various initial
parameter combinations that correspond roughly to parameter space explored in each column.
The first column shows a direct comparison between HD 189733 (dark gray line) and the
templates (orange lines). Column 2 shows the case of high $A_{fac}$ and low $q_{fac}$, i.e.,
weak facular emission but distributed across much of the stellar disk. Column 3 shows 
slightly stronger $q_{fac}$ but reduced $A_{fac}$. Finally, column 4 shows large $q_{fac}$
and small $A_{fac}$. We note that the HD 189733 spectrum is also plotted in columns 2-4
but is obscured by the model fits. 

For all three templates, a crucial trend is seen: as $q_{fac}$ increases, $A_{fac}$ must decrease
for a given photospheric spectrum. This is not surprising but has important consequences for
which parameter combinations in the contrast model are likely to be representative of
HD 189733's active surface. For example, the very low $Q$ and very high $q_{fac}$ case
shown in \autoref{fig:actvery} is unlikely to accurately represent the stellar surface since the
same parameter combination would drastically overestimate the line core emission in \autoref{fig:hacorefill}.
However, the less active case in \autoref{fig:actless} seems plausible: similar values of $q_{fac}$
and $Q$, or a facular coverage fraction of $\sim$8\%, are seen in the low $A_{fac}$, high $q_{fac}$
case for HD 87883 in \autoref{fig:hacorefill}.

One important caveat to the template comparison is that the
template spectra themselves contain some contribution from active regions. Thus we are not comparing
pure photospheres to HD 189733's active surface. However, this does not affect the
relationship between $q_{fac}$ and $A_{fac}$; one must decrease as the other increases
for a specific template spectrum, pure photosphere or not.
On the other hand, the \textit{value} of the parameters in \autoref{fig:hacorefill} may be
affected. For example, instead of $A_{fac}=0.08$ and $q_{fac}=4.06$ in the lower right
panel of \autoref{fig:hacorefill}, one or both values would need to be larger if HD 87883's 
spectrum contained no contributions from active regions. Thus the parameter
combinations shown in \autoref{fig:hacorefill} are likely lower limits to the true values.

\subsection{Discussion and summary of contrast results}

We have presented transit models for a planet with no atmosphere in order to explore
the contrast effect in H$\alpha$ that is produced when in-transit spectra, for which a portion of the
stellar disk is occulted, are compared with out-of-transit spectra. We find the following:

\begin{itemize}

\item Spots and filaments, for the physically reasonable surface coverage fractions and spot/filament parameters 
explored here, are unimportant in the H$\alpha$ contrast spectrum. The main contribution to the 
contrast effect comes from strong facular or plage emission.

\item Transits of the photosphere do not produce H$\alpha$ contrast in absorption; active regions
that include strong faculae/plage emission must be transited to produce $S_T$ in
absorption.

\item The facular coverage fraction must be $\gtrsim 5-10$\% and these facular regions must
be concentrated around the transit chord in order for the strongest observed H$\alpha$ 
transits to be reproduced. Large coverage fractions, no matter the value of $q_{fac}$, cannot
reproduce the observed H$\alpha$ line profiles if the distribution is uniform across the stellar
disk. This holds true even for very large coverage fractions of $>50$\% since the contrast
spectrum then begins to be weighted back towards the active region spectra, producing
emission instead of absorption.

\item Certain configurations of facular regions combined with values of $q_{fac} \sim 4.0$ are
able to produce relatively uniform transits with depths of $W_{H\alpha} \approx 0.007 - 0.015$,
similar to the observed transits shown in \autoref{fig:hatransits}. 

\item The comparison of less active template stars to HD 189733 suggests that for
$q_{fac} \sim 4$ the facular coverage fraction must be $\gtrsim 5-10$\%. This is similar
to what is needed to reproduce the observed transits, although these are lower limits
since the templates do not represent pure stellar photospheres.

\end{itemize}

While our simulations show that the contrast effect is able to reproduce the observed
in-transit $W_{H\alpha}$ values, it is unclear if the necessary parameters are physically
realistic. We believe facular coverage fractions of $\sim$10\%, and possibly higher, 
are not unreasonable considering the stronger activity level of HD 189733 compared
with the Sun. Solar facular coverage can reach $\sim$5\% during solar active periods \citep{shapiro15}
and plage coverage can reach $\sim$8\% \citep{foukal98}.
On the other hand, \citet{shapiro15} \citep[see also][]{foukal98,lockwood07} find that photometric
variations for more active stars are spot-dominated and that the ratio of facular/plage
coverage to spot coverage \textit{decreases} with increasing activity level. \citet{lanza11} find
that RV modulations in HD 189733's spectrum are best modeled with values of $Q=A_{fac}/A_{spot}$$\sim$0, i.e.,
they find no evidence of facular/plage effects on the measured RV values. Furthermore, 
\citet{lanza11} find spot filling factors of $\sim$1\% are able to reproduce HD 189733's photometric
variations across $\sim$4 weeks of nightly observations. Similar spot coverages are
found by \citet{herrero16} for the same data set. This would suggest values of $Q=0.1$
for the contrast model in order to produce $\sim$8\% facular/plage coverage.

The larger question for H$\alpha$ contrast models is what constitutes a reasonable
value of $q_{fac}$. We have shown that $q_{fac} \gtrsim 4.0$ is needed in order to
produce values of $W_{H\alpha}$ similar to what is observed. If this is the case,
then the normal or quiescent facular/plage regions on HD 189733 have H$\alpha$ 
emission line strengths similar to moderate flaring regions on the Sun.
Magnetic fields play an important role in heating the chromosphere and producing 
bright regions in the atmosphere \citep[e.g.,][]{hansteen07}. Indeed, simulations of
H$\alpha$ brightness in solar active regions show that the brightest regions correspond to
the strongest local magnetic field strengths \citep{leenaarts12}. HD 189733 has
a much stronger global magnetic field than the sun, with radial field values reaching
30-40 G across much of the stellar surface \citep{fares10}. Thus the larger field
values, and consequently heating rates, could have a significant effect on the emission strength of H$\alpha$ in
active regions.  

A final consideration is the requirement that the facular/plage regions be concentrated
very close to the planetary transit chord, or a latitude very near $+40^\circ$. Active latitudes are
known features of the solar surface \citep[e.g.,][]{vecchio12} and have been inferred via 
photometric transits for other stars as well \citep{sanchis11a,kirk16}. \citet{lanza11}
find that spots near $\approx 30^\circ-40^\circ$ are able to best reproduce the measured
RV variations. Thus it is not unreasonable that the facular/plage regions are located
near the transit chord. The specificity required, however, to reproduce any of the observed
$W_{H\alpha}$ transits is concerning, especially if active latitudes migrate as a function
of time. 

Although we cannot conclusively distinguish between models in which the observed $W_{H\alpha}$ transits are due to
absorption by the planet or from the contrast effect, we believe the parameter values required  to
reproduce the observations are rather specific and likely do not represent the average stellar disk
of HD 189733. Furthermore, the contrast effect cannot explain the $W_{H\alpha}$ values seen in
absorption immediately before the 2013 July 4 transit, before and after the 2015 August 4 transit,
and immediately after the 2006 August 21 transit, i.e., these extended transits are suggestive of
absorbing circumplanetary material. In addition, we demonstrated in \citet{cauley17} that
these pre- and post-transit signatures are abnormal and must be related to the planetary transit,
further making the case for a circumplanetary origin.
More detailed modeling of H$\alpha$ spectral line profiles in
faculae and plages for active stars are needed to determine if the large core strengths used here
are realistic. For now, however, we favor the planetary interpretation. 

\section{H$\alpha$ velocity measurements and models of planetary rotation}
\label{sec:velocities}

\citet{barnes16} presented an analysis of the same three HARPS transits presented in this work. They
note a trend in the velocity centroids of the H$\alpha$ transmission spectra, when calculated in the frame
of the planet, moving from red-shifted to blue-shifted (their Figure 3). This is cited as evidence that
the absorption may not arise in the planetary atmosphere since the absorption line profiles
should be centered at zero velocity in the planetary reference frame. In this section we present 
measurements of the H$\alpha$ transmission spectrum velocity centroids and models 
of atmospheric absorption which include planetary rotation. We demonstrate that velocities in the
upper atmosphere of HD 189733 b might explain the in-transit velocity trends identified by \citet{barnes16}.

To date, only a handful of studies have presented observational signatures of atmospheric dynamics
for massive exoplanets. The first detection of a day-to-night side wind, a result consistent with
predictions by a variety of detailed hot Jupiter circulation models
\citep[e.g.,][]{showman02,showman09,rauscher10,menou10,rauscher14}, was made by \cite{snellen10} for
HD 209458 b. They observed a $\sim$2 km s$^{-1}$ blue-shifted offset in a CO absorption signal
which matches closely with the magnitude of the flow velocity from atmospheric models
\citep[e.g.,][]{showman08}. \citet{snellen14} measured an equatorial rotational velocity of 25 km
s$^{-1}$ for the young massive planet $\beta$ Pic b using cross-correlated thermal signatures of CO
and H$_2$O. \citet{wyttenbach} measured an 8$\pm$2 km s$^{-1}$ blue-shift in the \ion{Na}{1} doublet
and suggested that the large velocity might be the result of high-altitude winds. 

Most recently, studies by \citet{louden} and \citet{brogi16} have demonstrated detections of the planetary
rotational velocity and day-to-night side winds of HD 189733 b.  For \ion{Na}{1}, \citet{louden}
also claim a detection of a spatially resolved, eastward super-rotating equatorial jet.
\citet{brogi16} search for a similar feature in their near-IR CO measurement but are unable to place
constraints on a jet velocity. Both studies find planetary rotational velocities that are
consistent with synchronous rotation and day-to-night side wind speeds of $\sim$2 km s$^{-1}$. The
results of \citet{brogi16} and \citet{louden} demonstrate the exciting possibility of measuring the
atmospheric dynamics of hot Jupiter atmospheres using ground-based, high-resolution spectra.

In \citet{cauley16} we presented an H$\alpha$ velocity measurement (see \autoref{eq:vhalpha} below)
and the corresponding in-transit $v_{H\alpha}$ values for two Keck HIRES transits.  \citet{barnes16}
investigated the H$\alpha$ line velocities of the three HARPS data sets examined in this paper. They
present velocity profiles as a function of orbital phase \citep[see Figure 3 of][]{barnes16} but do
not calculate velocities for individual absorption line profiles. They reference visual features in
the velocity maps as evidence of trends in the line velocities across the transit. Here we present
explicit velocity measurements of the individual HARPS and Keck spectra. As we show below, the
individual HARPS transmission spectra are fairly noisy and result in highly uncertain values of
$v_{H\alpha}$. The Keck velocities are more tightly constrained due to the much higher
signal-to-noise of the H$\alpha$ spectra. 

We perform all of our velocity analysis in the stellar rest frame. After correcting the
observed spectra for the system radial velocity and the Earth's heliocentric motion, this is also
the frame of the observer. Shifting the transmission spectra by the planetary radial velocity can
confuse and mask atmospheric velocities that may be contributing to the line profile, as we demonstrate
below.

\begin{figure}[htbp]
   \centering
   \includegraphics[scale=.65,clip,trim=35mm 12mm 25mm 0mm,angle=0]{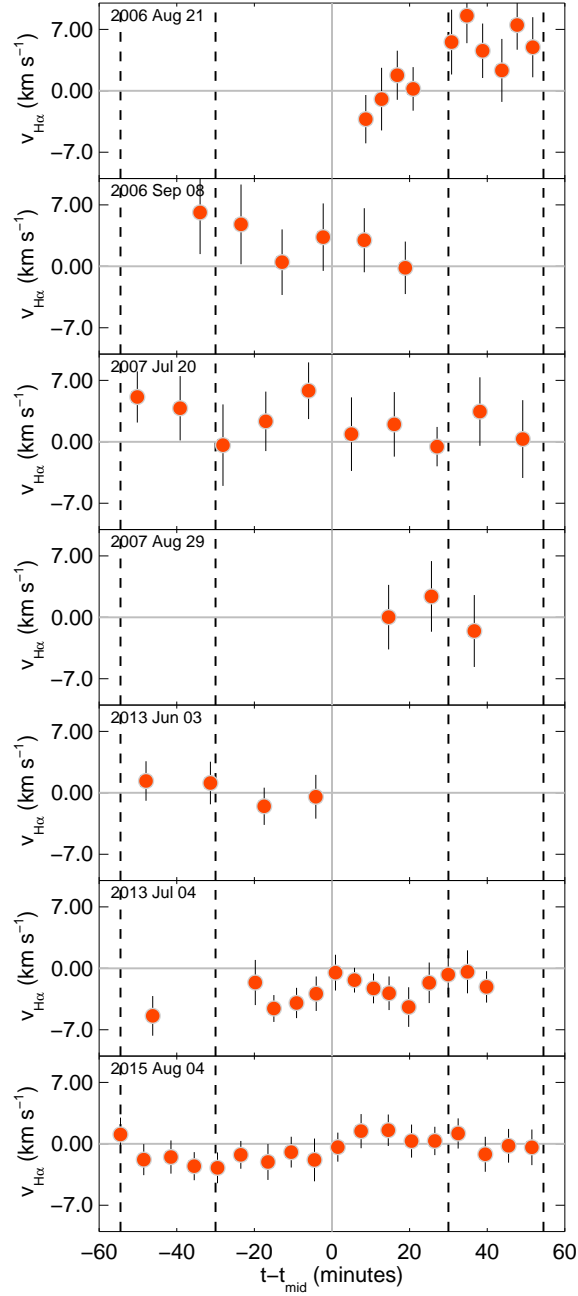}
   \figcaption{Values of $v_{H\alpha}$ from \autoref{eq:vhalpha} for all in-transit points showing $1\sigma$ absorption.
   The uncertainties for most of the HARPS measurements are $\sim$3-4 km s$^{-1}$ while the Keck uncertainties
   are $\sim$1-2 km s$^{-1}$. The vertical dashed lines mark the transit contact point. 
   There are no clear patterns in the velocity measurements for any of
   the transits except 2015 August 4. \label{fig:vhaall}}
\end{figure}

The line velocity index from \citet{cauley16} is defined as: 

\begin{equation}\label{eq:vhalpha}
v_{H\alpha} = \frac{\sum\limits_{v=-40}^{+40} v \left(1-F_{v}\right)^2}{\sum\limits_{v=-40}^{+40} \left(1-F_{v}\right)^2}
\end{equation}

\noindent where $F_v$ is the transmission spectrum flux at velocity $v$. The index is essentially
the velocity vector $v$ weighted by the square of the transmission spectrum flux. The square
of the flux is chosen to provide stronger weight to deeper portions of the transmission spectrum. We only calculate
$v_{H\alpha}$ for observations that show $\geq 1 \sigma$ absorption. The uncertainty on the
individual $v_{H\alpha}$ points is calculated by taking the standard deviation of the mean of
velocities corresponding to the $(1-F_v)^2$ values that comprise 68\% of the total weight. This
produces larger uncertainties for lines where the weights are comparable at many different
velocities, which is the case for the noisy HARPS $S_T$ profiles.  The measured $v_{H\alpha}$ values
are shown in \autoref{fig:vhaall}. The typical HARPS uncertainties are $\sim$3-4 km s$^{-1}$ while
the Keck uncertainties are $\sim$1-2 km s$^{-1}$, although they are slightly larger for the 2006
August 21 transit. 

It is clear from \autoref{fig:vhaall} that there is no obvious $v_{H\alpha}$ trend across epochs,
even for the transits that show consistent $W_{H\alpha}$ values in absorption (e.g., 2007 July 20,
2013 July 4, and 2015 August 4). The only date that shows any transit pattern is 2015 August
4: the velocities change from slightly blue-shifted to slightly red-shifted from the first half of
the transit to the second, although the mean of $v_{H\alpha}$ for each half of the transit
is only different from zero at the $\sim 1-2 \sigma$ level. The 2013 July 4 transit shows a mild blue-shifted offset of $\sim$2 km
s$^{-1}$ but $v_{H\alpha}$ values from the first half of the transit are plagued by weak
$W_{H\alpha}$ values. Little information is gleaned from the partial transits. We will explore the
2015 August 4 trend using models of planetary rotation and velocities as a product of the contrast
effect from \autoref{sec:contrast}.

\subsection{Planetary rotation models}

In order to explore physical scenarios for the H$\alpha$ line velocities presented above, we have
simulated transmission spectra through a rotating planetary atmosphere. We use the same stellar and
planetary parameters that were presented in \autoref{sec:contrast} and the same PHOENIX model
spectra are used as the intrinsic stellar spectra. We neglect active regions on the stellar
surface and assume a pure photosphere (see \autoref{sec:convels}). The stellar radial velocity as a function of in-transit time 
is included in the line profile calculations. 

The planetary atmosphere is assumed to be of uniform density.
This choice is motivated by the models of \citet{christie} who found that the number density of
$n=2$ hydrogen was approximately constant across 3-4 orders of magnitude in atmospheric pressure.
Assuming a hydrostatic versus uniform density atmosphere has little effect on the simulations. The
planetary H$\alpha$ absorption profile is approximated as a velocity-broadened delta function
where the broadening parameter is labeled $b$ \citep[][]{draine,cauley15,cauley16}.
The planet is assumed to rotate rigidly throughout the atmosphere. 

We have also investigated the effects of superrotating equatorial jets \citep{showman08,rauscher10} 
and find that they produce similar effects as rigid rotation, although the
jet velocities required to produce the same line profiles are larger than the rotational velocities
since the jets occupy a smaller portion of the atmosphere. We do not include jets explicitly in
the rest of the model discussion but we will discuss them as part of the general effect
of large rotational velocities.

\begin{figure*}[htbp]
   \centering
   \includegraphics[scale=.70,clip,trim=10mm 51mm 15mm 60mm,angle=0]{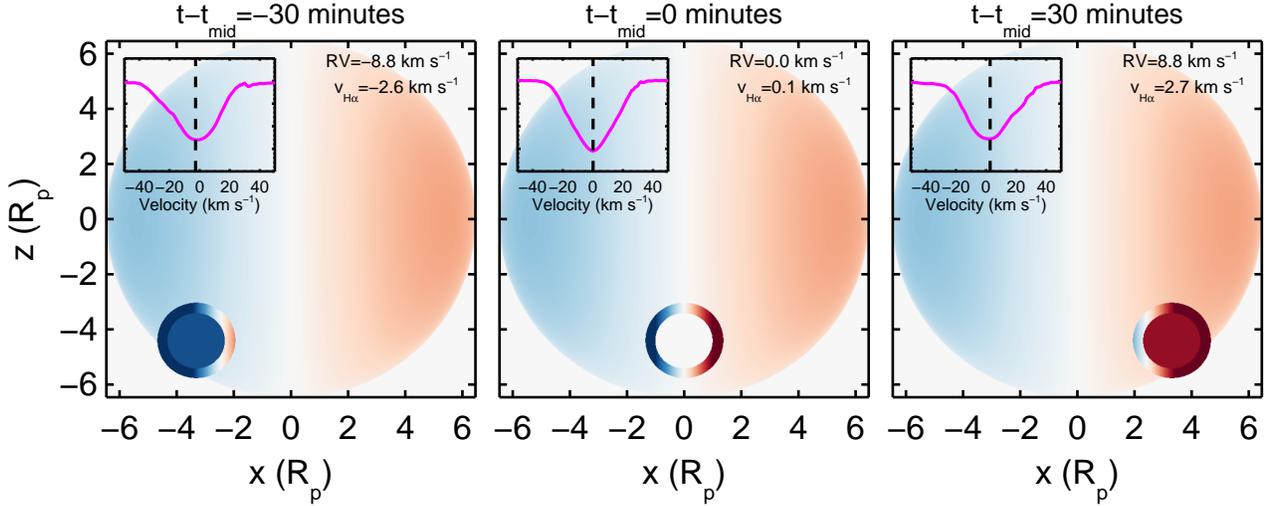}
   \figcaption{Example of how planetary rotation affects the measured transmission spectrum. The colors
   are representative of the radial velocity of the colored portion. Darker colors represent larger velocities.
   The stellar rotational velocities are weighted by the limb darkened intensity to visually represent
   the weighted velocity contribution of the stellar disk. The bulk motion of the planetary disk is given a single color and is 
   given in the upper-right of each panel. The measured $v_{H\alpha}$ from \autoref{eq:vhalpha} for
   the inlaid magenta transmission spectrum is also given in the upper-right corner. The $v_{H\alpha}$
   values are much lower than the planetary RV for times when the planet is near the limb, suggesting
   that any planetary rotation will reduce the measured line velocities. This
   example shows absorption lines for an atmosphere with $T_{exo}=7800$, $v_{rot}=10$ km s$^{-1}$,
   $b=4.0$ km s$^{-1}$, and $\rho=3.0\times10^{-23}$ g cm$^{-3}$. Note that the individual $v_{H\alpha}$ values do not
   change much for different values of $T_{exo}$, $b$, and $\rho$; it is dominated by $v_{rot}$. \label{fig:veltran}}
\end{figure*}

\begin{figure*}[htbp]
   \centering
   \includegraphics[scale=.70,clip,trim=5mm 30mm 25mm 40mm,angle=0]{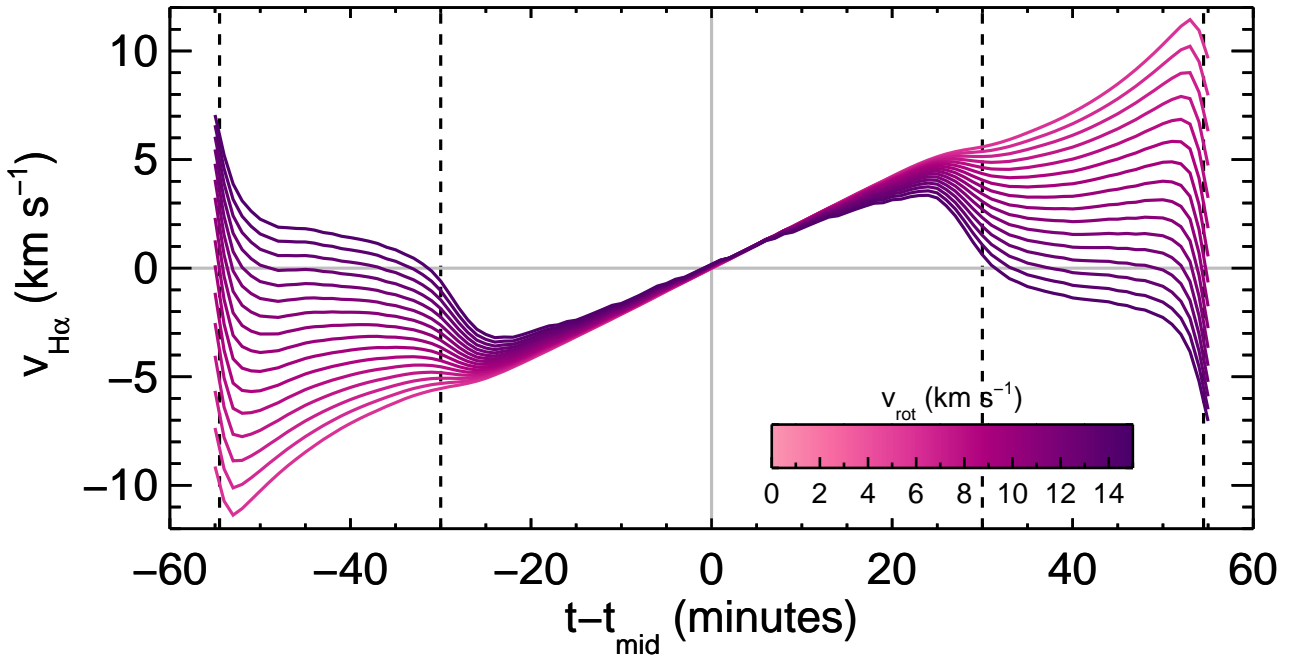}
   \figcaption{H$\alpha$ transmission spectrum velocity measurements, defined by \autoref{eq:vhalpha},
   as a function of time from mid-transit. The color indicates the equatorial planetary rotational
   velocity, the values of which are specified in the inset color bar. The transit contact points are
   marked with vertical dashed lines. As the planet rotates faster,
   absorption lines in the planetary atmosphere are shifted further away from the planetary
   orbital velocity upon ingress and egress, creating the slope changes in $v_{H\alpha}$ seen near
   $\pm 25$ minutes. Even the case of no rotation shows a depressed line velocity compared
   with the orbital velocity. This is due to CLVs described in \autoref{sec:contrast}. This set of models
   was calculated with the same atmospheric parameters used in \autoref{fig:veltran}. \label{fig:velmods}}
\end{figure*}

\autoref{fig:veltran} shows the effect of a rotating atmosphere on the transmission spectrum for a
prograde planetary orbit. Upon ingress (left panel) the portion of the planet's atmosphere dominating
the transmission spectrum is moving \textit{away} from the observer while the planet's bulk
motion is \textit{towards} the observer. The net effect is to produce a measured centroid velocity,
or in our case $v_{H\alpha}$, that is significantly less than the bulk planetary velocity. This was 
first demonstrated by \citet{kempton12} (their Figure 8) and reiterated in \citet{louden} (their Figure 2). 
The same effect is seen upon egress (right panel).

Simulated $v_{H\alpha}$ curves for the entire transit are shown in \autoref{fig:velmods} where
the effect of higher rotational velocities can be seen in the suppression of $v_{H\alpha}$
relative to the line-of-sight orbital velocity upon ingress and egress. For the highest rotational
velocities, the atmosphere is moving fast enough to cause $v_{H\alpha}$ to have the opposite
sign compared to the planet's bulk motion. We note that for strong atomic lines even the case of
no rotation produces suppressed line velocities due to the CLVs discussed in \autoref{sec:contrast}. 
For transits that are sampled asymmetrically in time, CLVs could produce velocity shifts
in the average transmission spectrum \citep[e.g., see][]{louden,wyttenbach}.

\autoref{fig:velmods} shows that for sufficiently large $v_{rot}$ the measured $v_{H\alpha}$ values
can be very small or even have the opposite sign relative to the planetary bulk velocity. While
these transit curves cannot explain the erratic velocities of the HARPS data or the 2006 August 21
Keck data, they may provide evidence for what we are seeing in the 2013 July 4 and 2015 August 4 
Keck transits and perhaps the first half of the 2013 June 3 transit. 

\begin{figure*}[htbp]
   \centering
   \includegraphics[scale=.70,clip,trim=10mm 10mm 10mm 40mm,angle=0]{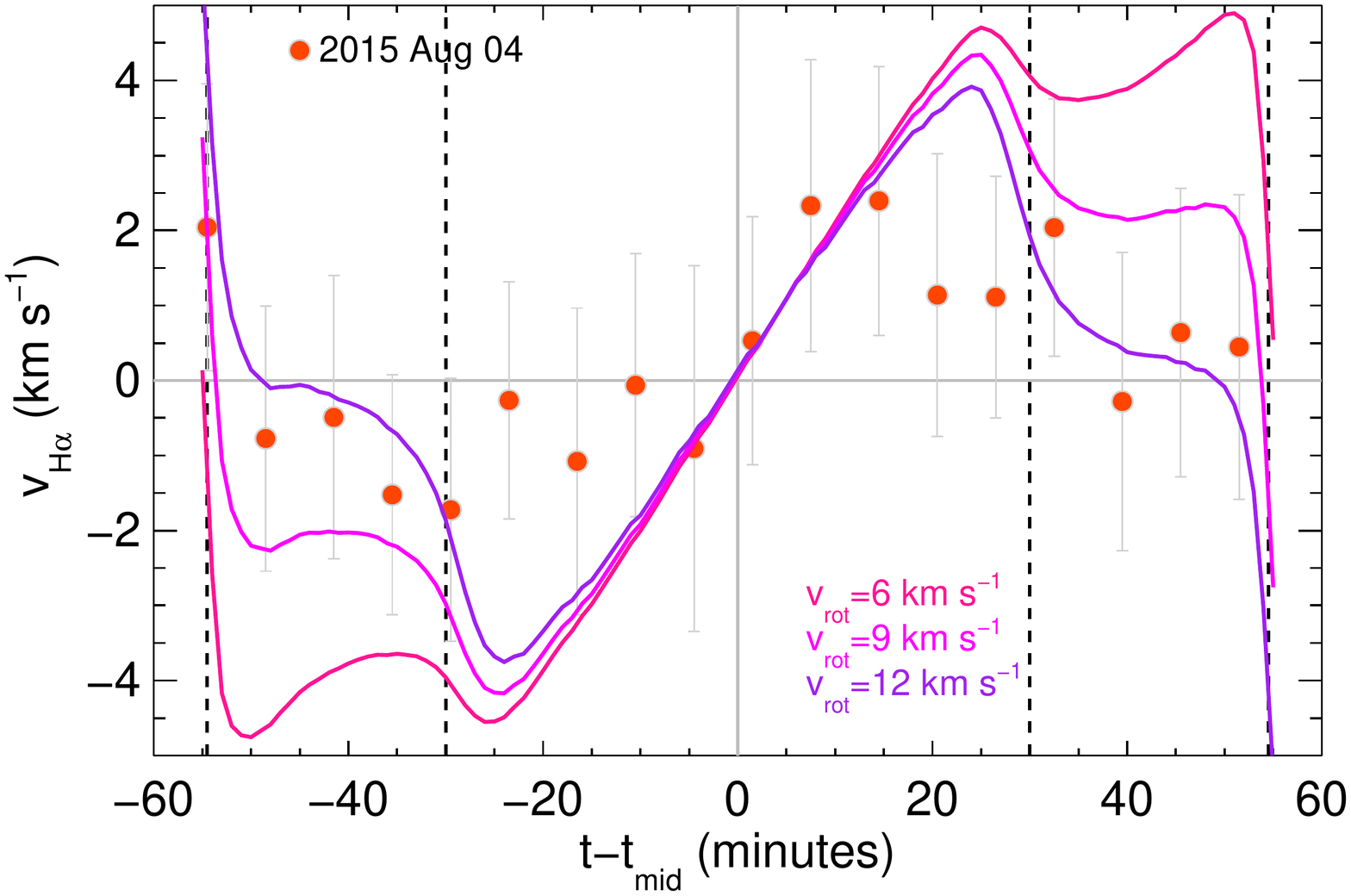}
   \figcaption{Comparison of planetary rotation models with the observed $v_{H\alpha}$ values
   from the 2015 August 4 transit. Three representative rotation models from \autoref{fig:velmods} are shown
   and the plot scale is reduced to show the detailed shape of the model curves. The models
   produce essentially identical $v_{H\alpha}$ values between second and third contact and do
   a poor job of matching the data for times immediately after second contact and immediately
   before third contact. However, the $v_{rot}=$9 and 12 km s$^{-1}$ models show similar
   $v_{H\alpha}$ values compared with the data between first and second and third and fourth
   contacts. 
    \label{fig:vmodcomp}}
\end{figure*}

\autoref{fig:vmodcomp} shows three different rotation models plotted against the 2015 August 4 $v_{H\alpha}$
values. Rotational velocities $\lesssim 9$ km s$^{-1}$ do not match the data well; velocities
between 10-12 km s$^{-1}$ are required to produce the small $v_{H\alpha}$ values between
first and second and third and fourth contact. Values of $v_{rot} \gtrsim 12$ km s$^{-1}$ begin
to produce $v_{H\alpha}$ values that are too small, i.e., of the wrong sign. The models do not
do a good job of reproducing $v_{H\alpha}$ near $-20$ and $+20$ minutes; all of the models
produce $v_{H\alpha}$ larger than the measured values. Overall, however, these models
demonstrate that large velocities in the extended atmosphere can produce the small observed
velocities. 

\subsection{Velocities from contrast models}
\label{sec:convels}

\begin{figure*}[htbp]
   \centering
   \includegraphics[scale=.70,clip,trim=10mm 40mm 5mm 80mm,angle=0]{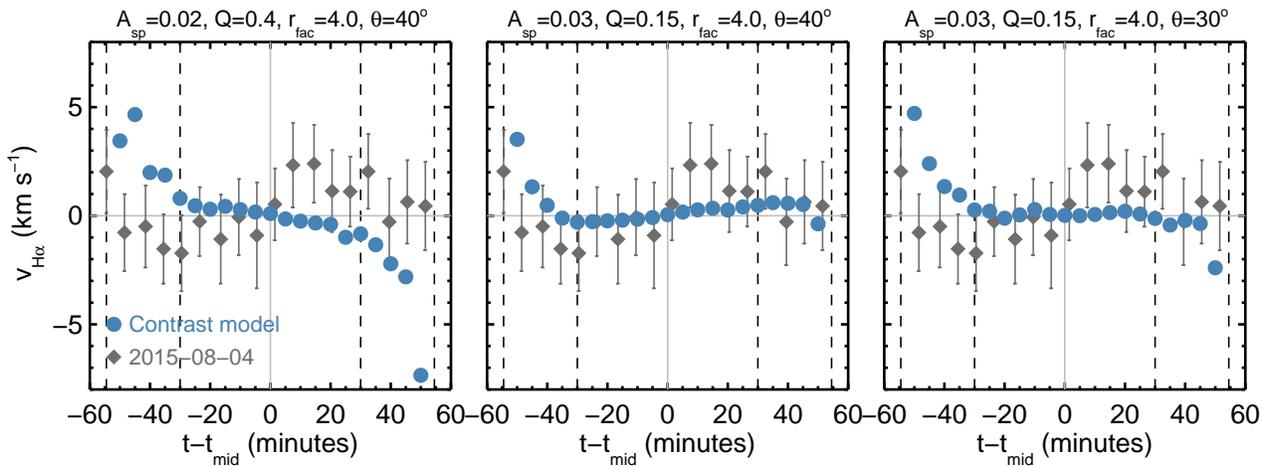}
   \figcaption{Velocities from the contrast models (blue circles) overplotted on the $v_{H\alpha}$
   values from the 2015 August 4 transit. The left panel corresponds to the contrast model in
   \autoref{fig:actless}, the middle panel to \autoref{fig:actvery}, and the right panel to \autoref{fig:actvery_30}.
   The contrast velocities from the middle panel do a fair job of reproducing the observed $v_{H\alpha}$
   values. Other contrast scenarios besides the very active case do not reproduce the observations
   well and in fact show the opposite trend, i.e., a transition from red-shifted to blue-shifted velocities 
   from the first half to the second half of the transit.
    \label{fig:convels}}
\end{figure*}

It is possible for certain contrast model scenarios to produce $v_{H\alpha}$ values similar to what is
observed for the 2013 July 4 and 2015 August 4 transits. We have calculated $v_{H\alpha}$ for
the contrast model line profiles. The $v_{H\alpha}$ values for the contrast models from \autoref{fig:actless}
(left panel), \autoref{fig:actvery} (middle panel), and \autoref{fig:actvery_30} (right panel) are
shown in \autoref{fig:convels}. The contrast model line profiles show trends similar to the
2015 August 4 data, although only the very active case (middle panel) shows the blue-shifted
to red-shifted pattern from the first half of the transit to the second half. Although not shown here,
all of the other cases explored in \autoref{sec:contrast} show something similar to the
left panel of \autoref{fig:convels}: large red-shifted velocities during the first half of the transit
and blue-shifted velocities during the second half. More active cases than the middle panel
tend to move the other direction: larger and larger blue-shifted velocities and then red-shifted
velocities during the second half. 

Overall, the velocities from the contrast models do not reproduce the observed 2015 August 4 velocities
except for the case of a very active stellar surface with the active regions centered very near the
planet's transit chord (see \autoref{fig:actvery}). For this reason, and those given concerning the
planetary rotation models, we do not believe the contrast velocities represent a convincing
explanation for the observations. However, further in-transit observations should be conducted
to strengthen or reject this argument.

\subsection{Discussion of the rotation models and H$\alpha$ velocities}

The rotation models presented here are meant to demonstrate that the low velocities, i.e., much less
than the in-transit planetary line-of-sight velocity, observed in the H$\alpha$ transits are not 
necessarily intrinsic to the star because of their magnitude. While very hot Jupiters such as 
HD 189733 b are generally assumed to be synchronously rotating, there are no measurements of
hot Jupiter rotation rates outside of the studies done by \citet{brogi16} and \citet{louden}. Both
\citet{brogi16} and \citet{louden} found evidence for a synchronously rotating HD 189733 b. However,
these measurements were made in CO\footnote{\citet{brogi16} measure the strongest signal in CO but
also marginally detect H$_2$O.} \citep{brogi16} and \ion{Na}{1} \citep{louden}. The CO measurements probe
much higher pressures deeper in the atmosphere than H$\alpha$ and \ion{Na}{1}, which probe
pressures of 10$^{-6}$-10$^{-9}$ bar \citep{christie,wyttenbach}. The \ion{Na}{1} measurements by \citet{louden} do not
examine the velocities of individual observations and instead fit the average in-transit line
profile. \citet{louden} also do not include differential limb darkening, opting to use a broadband
limb darkening law. This could significantly affect the velocities of the model line profile
\citep{czesla15} and should be investigated. 

Even if a hot Jupiter is synchronously rotating, velocities larger than the equatorial rotational
velocity may be present in the extended atmosphere. \citet{rauscher14} find wind speeds of
$\sim$7-11 km s$^{-1}$ in the upper atmospheres of both a synchronously and a quickly rotating
HD 189733 b. Their models were calculated at pressures of 10$^0$-10$^{-6}$ bar, which begins
to probe the potential H$\alpha$ formation region. Thus it is not implausible that even stronger
winds may form at higher altitudes or that unexplored atmospheric dynamics are contributing to the  
H$\alpha$ transmission spectra. 

While the most active contrast model is able to produce $v_{H\alpha}$ values similar to what is
observed for the 2015 August 4 transit, we do not believe this is the correct model due to
required specificity of the model parameters. In other words, matching the velocities 
requires all of the most active contrast model parameters to be accurate whereas matching only 
the $W_{H\alpha}$ observations provides more leeway in the active region coverage
fraction and facular emission line strength. On the other hand, the 2015 August 4 transit
does seem to be unique in that it shows the largest $W_{H\alpha}$ signal and the strongest
out-of-transit \ion{Ca}{2} emission. Although it seems unlikely, we cannot definitively rule out the most
active case as an explanation for the both $W_{H\alpha}$ and $v_{H\alpha}$. 

We emphasize that the H$\alpha$ transits do not show a consistent velocity signal across
epochs and as a result no firm conclusions can be reached concerning their origin. The individual 
HARPS spectra are especially noisy and the velocity uncertainties derived here are large making
them even less useful for understanding the in-transit signal. Further short-cadence
H$\alpha$ transit observations are needed to work towards clarifying the velocity signal
and, by extension, the in-transit H$\alpha$ absorption line profiles. 

\section{Summary and conclusions}
\label{sec:conclusion}

We have presented an analysis of H$\alpha$ transmission spectra for 7 transits of HD 189733 b
that span almost a decade. Five of these H$\alpha$ transits are from archival HARPS
and Keck data, while two were previously analyzed by our group in \citet{cauley15,cauley16}.
Four of these transits show significant and consistent H$\alpha$ spectra in
absorption throughout the transit while three others show strong variations that may be due
to changes in stellar activity. The irregular changes may also be due to transiting gas not 
bound to the planet. Our main conclusions are the following:

\begin{itemize}

\item We do not find evidence of a clear relationship between the stellar activity level and the
strength of the in-transit H$\alpha$ signal. The
outlier in this case is the data from 2015 August 4 which shows both the strongest H$\alpha$
absorption and the highest stellar activity level. If the absorption signal arises in the planetary atmosphere,
this reveals a potential stellar activity threshold below which the ionizing flux from the star is 
not high enough to create large H$\alpha$ transit depths.
This should be investigated further with future transits and simultaneous UV measurements 
of the stellar activity level.

\item We explored detailed simulations of the H$\alpha$ contrast effect for HD 189733 b. We find that
large facular/plage coverage fractions of $\gtrsim 5-10$\% and very strong facular/plage
core emission strengths of $q_{fac}\sim 4$ are required to reproduce the observed H$\alpha$
observations. Furthermore, these facular/plage regions must be concentrated very close
to the transit chord of the planet. Spots and filaments do not have a significant effect
on the H$\alpha$ contrast spectrum. 

\item Due to the specificity of the contrast parameters required
to reproduce the measured $W_{H\alpha}$ values, combined with the natural explanation
of absorption in the thermosphere \citep{christie}, we favor a planetary atmospheric
origin for the H$\alpha$ transmission spectra. A similar argument can be made for
interpreting the $v_{H\alpha}$ measurements. However, detailed models of active regions
on active stars such as HD 189733 are needed to understand if the necessary contrast
parameters are reasonable. 

\item We have also explored the velocity centroids of the measured H$\alpha$ transmission
spectra using models of planetary rotation. We find that planetary rotational velocities of $\sim$9-12 km s$^{-1}$
are able to produce in-transit $v_{H\alpha}$ values similar to those from the 2015 August 4 
transit. These large velocities are representative of dynamics in the very extended atmosphere
and do not necessarily suggest that the planet is rotating faster than the synchronous rate.
These models demonstrate that large atmospheric velocities can produce the small
observed $v_{H\alpha}$ values and that they do not need to originate on the stellar
surface. However, there is no consistent velocity pattern across epochs so the results
of the rotation models cannot be broadly applied. Further short-cadence transits
at very high S/N are needed to test these ideas.

\end{itemize}

Future and current planet hunting missions, such as the \textit{Transiting Exoplanet Survey Satellite} (\textit{TESS})
and the ground-based Kilodegree Extremely Little Telescope \citep[KELT][]{pepper07} and the Multi-site All-Sky CAmeRA \citep[MASCARA][]{talens17},
have and will detect many hot planets transiting relatively bright stars. 
This will significantly increase the number of objects for which short-cadence, high-resolution 
optical transmission spectroscopy can be performed with 4-meter or 10-meter class telescopes. 
In addition, thirty-meter class telescopes will be able to perform similar observations on much fainter systems.
Stars with high activity levels should be targeted as comparison cases with HD 189733 b. Indeed,
active stars may be the only systems with hot planets exhibiting H$\alpha$ absorption in
their extended atmospheres \citep{christie}. These systems will also act as testbeds for disentangling the
contrast effect from true planetary absorption.

Ground based H$\alpha$ observations of the extended atmospheres of hot planets offers a complimentary alternative
to the exospheric \textit{Hubble Space Telescope} (\textit{HST}) Ly$\alpha$ 
observations \citep{vidal,desetangs,desetangs12,bourrier13,vidal13,kulow,ehren15}. While the H$\alpha$
observations do not probe the escaping exosphere, this ground-based
approach may become the best option for measuring the base of evaporative flows
due to the impending loss of \textit{HST} and its spectroscopic UV capabilities. It is also
critical to develop ground based transmission spectrum programs in order to plan for and complement
future space missions, such as the \textit{Large UV/Optical/IR Surveyor} (\textit{LUVOIR}). HD 189733
is a benchmark for testing the usefulness of the H$\alpha$ diagnostic and
the relationship between stellar activity and the planetary thermosphere. It is thus
important to further investigate the observed HD 189733 signals, along with any future detections,
to determine the origin of the H$\alpha$ signal and strengthen or repudiate the arguments presented here.

\bigskip

{\bf Acknowledgments:} We thank the referee for their critique of this manuscript, which helped
improve the clarity and style. A portion of the data presented herein were obtained at the W.M. Keck
Observatory from telescope time allocated to the National Aeronautics and Space Administration
through the agency's scientific partnership with the California Institute of Technology and the
University of California. This work was supported by a NASA Keck PI Data Award, administered by the
NASA Exoplanet Science Institute. The Observatory was made possible by the generous financial
support of the W.M. Keck Foundation. The authors wish to recognize and acknowledge the very
significant cultural role and reverence that the summit of Mauna Kea has always had within the
indigenous Hawaiian community. We are most fortunate to have the opportunity to conduct
observations from this mountain. A. G. J. is supported by NASA Exoplanet Research Program 
grant 14-XRP14 2- 0090 to the University of Nebraska-Kearney.
This work was completed with support from the National Science
Foundation through Astronomy and Astrophysics Research Grant AST-1313268 (PI: S.R.). This work has
made use of NASA's Astrophysics Data System.

\end{document}